\documentclass[prb,preprint]{revtex4-1} 

\usepackage{amsmath}  
\usepackage{amsfonts} 
\usepackage{graphicx} 

\begin{document}


\title{Light deflection around a spherical charged black hole to second order.  Multivariate Pad\'e approximants }


\author{Carlos Mar\'in}
\email{cmarin@usfq.edu.ec}
\affiliation{Department of Physics, Universidad San Francisco de Quito, Diego de Robles y V\'ia Interoce\'anica }

\author{Jorge Poveda }
\email{jorge.okuden@gmail.com} 
\affiliation{Department of Physics, Universidad San Francisco de Quito, Diego de Robles y V\'ia Interoce\'anica} 



\date{\today}
\begin{abstract}{From the Reissner-Nordstr\"om metric we  obtain the higher-order terms for the deflection of light around a massive-charged black hole using the Lindstedt-Poincar\'e method to solve  the equation of motion of a photon around the compact object. The corrections are performed developing the expansion in terms of $\epsilon=\frac{r_{c}}{b}$  and $\frac{\delta}{b}= \frac{Q^{2}}{6 \pi \epsilon_{0}Mc^{2} b}$.  Additionally, from the perturbation expansion, we obtain multivariate Pad\'e approximants. We also show how these are a better fit for the numerical data than the original formal Taylor series.} 
\end{abstract}


\maketitle 
\section{Introduction}
\label{intro}

One of the most interesting predictions of the "General Theory of Relativity" (GTR) is the existence of black holes, regions of space-time from which despising quantum effects, you can not escape. The Physics of a massive spherical black hole with no electric charge and rotational movement is described by the Schwarzschild metric, introduced by Karl Schwarzschild in 1916.
 For a spherical symmetric space-time with a mass $M$, the invariant interval is \cite{Kenyon,Misner,Marin2}:
\begin{eqnarray}
\left(ds\right)^{2} = \gamma\left(c dt\right)^{2} - \gamma^{-1}\left(dr\right)^{2} -r^{2}\left(d\Omega \right)^{2}  \label{eq:MetricSchwar}
\end{eqnarray}
where $\left( d \Omega \right)^{2} = \left(d \theta\right)^{2} +sin^{2}\theta \left(d \phi\right)^{2}$ , with coordinates  $x^{0}=ct$, $x^{1}=r$, $ x^{2}=\theta$ 
and  $x^{3}=\phi$. $\gamma = 1 - \frac{r_{s}}{r}$ ,being $ r_{s} = \frac{2GM}{c^{2}}$ the Schwarzschild radius. The arc length  $s$ satisfies the relation $ds^{2} = g_{\mu \nu}dx^{\mu}dx^{\nu}$, then, the covariant metric tensor is:

\begin{equation}
(g_{\mu\nu})=\left(\begin{array}{cccc}
\gamma & 0 & 0 & 0\\
0 & -\gamma^{-1} & 0 & 0\\
0 & 0 & -r^{2} & 0\\
0 & 0 & 0 & -r^{2}\sin^{2}\theta
\end{array}\right).
\end{equation}  \\
 
We have two singularities in this metric. The first when $\gamma=0$ or $r= r_{s}$ is a mathematical singularity that can be removed  by a convenient coordinate transformation like the one  introduced by Eddington in 1924 or Finkelstein in1958
\cite{Kenyon,Marin2}:

\begin{equation}
\hat{t}=t\pm\frac{r_{s}}{c}ln\left|\frac{r}{r_{s}}-1\right|.  \label{eq:Eddington}
\end{equation}

With this coordinate transformation, the invariant interval can be written as:

\begin{eqnarray}
\left(ds\right)^{2} = c^{2}\left(1 - \frac{r_{s}}{r}\right)\left(d \hat{t} \right)^{2} - 
\left(1 + \frac{r_{s}}{r}\right) \left(dr\right)^{2} \mp 2c \left(\frac{r_{s}}{r}\right) d\hat{t} dr
- r^{2} \left(d \Omega\right)^{2}.
\end{eqnarray}

The first transformation in equation (\ref{eq:Eddington}) 
\begin{equation}
\hat{t}=t + \frac{r_{s}}{c}ln\left|\frac{r}{r_{s}}-1\right|  \label{eq:Eddington1}
\end{equation}

describes a black hole, while the second one:

\begin{equation}
\hat{t}=t - \frac{r_{s}}{c}ln\left|\frac{r}{r_{s}}-1\right|  \label{eq:Eddington2}
\end{equation}
represents what physicists call a "white hole" emitting material from a singularity in $r=0$ toward space-time.

The other singularity $r=0$ is physical, so it can not be removed. In this singularity, all known physical laws fail, and the curvature of space-time is infinite. If one particle reaches the event horizon ($r=r_s$), it will eventually falls to the singularity $r=0$, and it will never escape from the black hole neglecting quantum effects like Hawking radiation.

There is good evidence of the existence of black holes; for example the X-ray source known as Cygnus X-1 \cite{Marin2,Ramesh}. Cygnus X-1 emits X-rays and radio waves in an irregular way that contrasts with the regular emissions that are observed in a binary pulsar such as the PSR 1913 + 16. Cygnus X-1 is the companion of a blue supergiant star (HDE 226868) located at a distance of 2.5 kiloparsecs (8150 light years) from Earth. The mass of the blue supergiant is approximately 20 times that of the Sun. The wavelengths of the spectral lines of the supergiant is too cold to emit X-rays. Material ripped from HDE 226868 forms an accretion disk around Cygnus X-1 that is releasing an immense amount of energy. This energy heats the accretion disk to a temperature at which large amounts of X-rays are emitted. In conclusion, HDE 226868 and Cygnus X-1 form an X-ray binary with an orbital period of 5.6 days.

The mass of Cygnus X-1 can be calculated from the mass function:

\begin{equation}
f \left(M \right) \equiv  \frac{\left(M_{X} \sin i \right)^{3}}{\left(M_{X}+M_{C}\right)^{2}}=\left(a \sin i\right)^{3}\frac{\left(\frac{2 \pi}{\tau}\right)^{2}}{G}
\end{equation}

where $M_{X}$  is the mass of the X-ray source, $M_{C}$ is the mass of the blue supergiant, $i$ is the angle between the direction of observation and the normal to the plane of the orbit, $a$ is the length of the semimajor axis and $\tau$  is the orbital period of the X-ray binary. The mass of the source is estimated at approximately 15-16 times that of the Sun, making it a good candidate for a black hole. Other examples of binary systems that can host black holes are LMC X-3 and A0620-00 whose masses are estimated at nine times that of the Sun.

At the core of many galaxies such as the M-87, the active galaxy Centaurus A, the galaxy MCG-6-30-15, etc., there is evidence of the presence of black holes with masses that would range between $10^{6}$ and $3 \times 10^{9}$ of times the mass of the Sun \cite{GaryHorowitz} (supermassive black holes). Observations on the wavelengths corresponding to radio waves and X rays support this statement. These supermassive black holes may have played a very important role in the formation of galaxies in the early universe.

At the core of our galaxy, the Milky Way, there are signs of a swarm of black holes surrounding a supermassive black hole called Sagittarius A * (SgrA *) with a mass of about 4 million of  times the mass of the Sun \cite{MJReid}.

One of the largest  black holes observed so far is in the heart of the quasar OJ287  \cite{Valtonen1,PauliPihajoki,Valtonen2} at a distance of $3.5 \times 10^{9}$  light years from Earth. It has a mass between $17 \times 10^{9}$ and  $18 \times 10^{9}$ times that of the Sun. A smaller black hole with a mass of $100$  million times that of the Sun orbits around that supermassive black hole in an oval orbit with a period of twelve years. The advance of the perihelion due to the movement of translation of the smaller hole around the greater one is of 39 degrees in each orbit \cite{Valtonen1}.

The Chandra X-ray telescope \cite{Marin2} (name given in honor of the great theoretical astrophysicist Subrahmanyan Chandrasekhar who died in 1995), in orbit around the Earth, has recently obtained evidence of the existence of black holes in intermediate mass, between five hundred and twenty thousand times the mass of the Sun.

In general, depending on factors such as the temperature on the surface and the plasma physics in the solar corona, a star has a small electric charge. So, the contribution of the electromagnetic field in the metric is very small. However, when the star collapses (as nuclear fuel has run out) to become a black hole, one would expect that the entire charge will be ejected out of the star. For this reason, the black hole formed should be electrically neutral.  However, there could be some other mechanism (which we do not know) by which some black holes were not electrically neutral. In any case, only observations made in the future will allow us to discard or not this possibility.

In a previous paper \cite{MarinRodriguez} using the Schwarzschild metric we have obtained the higher-order terms for the deflection of light around a massive object (a black hole) using the Lindstedt-Poincar\'e method as well as  diagonal Pad\'e approximants from the perturbation expansion. In this paper,  we obtain from the Reissner-Nordstrom metric, the second-order terms for the deflection of light around a massive-charged black hole using the Lindstedt-Poincar\'e method to solve  the equation of motion of a photon around the compact object. The corrections are performed developing the expansion in terms of $\epsilon=\frac{r_{c}}{b}$  and $\frac{\delta}{b}= \frac{Q^{2}}{6 \pi \epsilon_{0}Mc^{2} b}$. Higher order terms can be obtained following  a procedure similar to the one we will illustrate in this paper. Also we obtain, in a way similar to the one we developed in the mentioned article \cite{MarinRodriguez}, multivariate Pad\'e approximants from the perturbation expansion.  We make use of  Pad\'e approximants on our asymptotic series for the deviation angle to increase its region of validity, and to improve as we shall see, matches the qualitative behavior of the deflection angle. 

It is worth mentioning that Pad\'e polynomials  were first  used in Cosmology with excellent results (see references \cite{Christine,Alejandro,Salvatore} for details).

\section{The Reissner-Nordstr\"om metric}
\label{ReissnerN}

The  exact solution to the Einstein field equations for a static space-time with spherical symmetry with charge $Q$ and a mass $M$ in the center of the coordinate system is given by the Reissner-Nordstrom metric. The invariant interval is \cite{Misner,tHooft}:
\begin{eqnarray}
\left(ds\right)^{2} = \Delta\left(c dt\right)^{2} - \Delta^{-1}\left(dr\right)^{2} -r^{2}\left(d\Omega \right)^{2}  \label{eq:MetricRN}
\end{eqnarray}
where  $\Delta \equiv 1- \frac{2GM}{rc^{2}}+\frac{Q^{2}G}{4\pi\epsilon_{0}r^{2}c^{4}}$ and 
$\left( d \Omega \right)^{2} = \left(d \theta\right)^{2} +sin^{2}\theta \left(d \phi\right)^{2}$ , with coordinates  $x^{0}=ct$, $x^{1}=r$, $ x^{2}=\theta$ 
and  $x^{3}=\phi$. 

Then, the covariant metric tensor is:

\begin{equation}
g_{\mu\nu}=\left(\begin{array}{cccc}
\Delta & 0 & 0 & 0\\
0 & -\Delta^{-1} & 0 & 0\\
0 & 0 & -r^{2} & 0\\
0 & 0 & 0 & -r^{2}\sin^{2}\theta
\end{array}\right).
\end{equation}  \\

The surfaces with infinite red-shift are obtained from $g_{00}=\Delta=0$. 
If $\frac{GM}{c^{2}} < \frac{Q }{c^{2}} \sqrt{\frac{G}{4\pi\epsilon_{0}}}$ we don't have such surfaces. In this case the singularity at $r=0$ is naked. However, this is forbidden by the hypothesis of the cosmic censorship \cite{Wald}. If $\frac{GM}{c^{2}} > \frac{Q }{c^{2}} \sqrt{\frac{G}{4\pi\epsilon_{0}}}$ we will have
two surfaces with infinite red-shift:

\begin{eqnarray}
r_{1}= \frac{GM}{c^{2}} - \left[\left(\frac{GM}{c^{2}}\right)^{2}-\frac{Q^{2}G}{4\pi\epsilon_{0}c^{4}}\right]^{\frac{1}{2}} \label{eq:redshift1}
\end{eqnarray}

\begin{eqnarray}
r_{2}= \frac{GM}{c^{2}} + \left[\left(\frac{GM}{c^{2}}\right)^{2}-\frac{Q^{2}G}{4\pi\epsilon_{0}c^{4}}\right]^{\frac{1}{2}} , \label{eq:redshift2}
\end{eqnarray}
where $\epsilon_{0}=8.854 \times 10^{-12}$ $\frac{Coul^{2}}{Nm^{2}} $ is the electric permitivity of the vacuum.
\\

If $\frac{GM}{c^{2}} = \frac{Q }{c^{2}} \sqrt{\frac{G}{4\pi\epsilon_{0}}}$ we will have only one of such surfaces:

\begin{eqnarray}
r_{\infty}= \frac{GM}{c^{2}} = \frac{Q }{c^{2}} \sqrt{\frac{G}{4\pi\epsilon_{0}}}
\label{eq:redshift3}
\end{eqnarray}
                
For a light ray the invariant interval is $ds=0$. In the equatorial plane   $\theta=\frac{\pi}{2}$, and then $d\theta =0$. Therefore we have:

\begin{eqnarray}
 \Delta\left(c dt\right)^{2} - \Delta^{-1}\left(dr\right)^{2} -r^{2}\left(d\phi \right)^{2} =0. 
\end{eqnarray}
or
\begin{eqnarray}
 \Delta c^{2} - \Delta^{-1}\left(\frac{dr}{dt}\right)^{2} -r^{2}\left(\frac{d\phi}{dt}\right)^{2} =0. 
\end{eqnarray}

From the last equation we can write:

\begin{eqnarray}
\left(\frac{dr}{dt}\right)^{2}=\Delta \left(\Delta c^{2}-r^{2} \left(\frac{d\phi}{dt}\right)^{2}\right)
\label{eq:horizons}
\end{eqnarray}

The event horizons are obtained setting $\frac{dr}{dt}=0$ for any $\theta, \phi$ and so $\Delta =0$. Then, the event horizons coincide with the surfaces with an infinite red-shift, being able to have two, one or no horizons of events.

Now, considering the equivalence between inertial and gravitational mass, the total energy of a spherical object (like a star)  with rest mass energy $M_{0}c^{2}$, charge $Q$ (uniformly distributed in the volume of the sphere)  and radius $R$ is:

\begin{eqnarray}
Mc^{2}=M_{0}c^{2}+\frac{3}{5}\frac{kQ^{2}}{R}-\frac{3}{5}\frac{GM^{2}}{R} ,
\label{eq:totalmassenergy}
\end{eqnarray}

where $k=\frac{1}{4\pi \epsilon_{0}}$.  The second and third terms in (\ref{eq:totalmassenergy}) represent the Coulomb energy and the gravitational binding energy, respectively. From this equation we can get:

\begin{eqnarray}
M=\frac{1}{2aG}\left(\left(R^{2}c^{4}+4aG\left(M_{0}Rc^{2}+akQ^{2}\right)\right)^{\frac{1}{2}}-Rc^{2}\right) ,
\label{eq:M}
\end{eqnarray}
\begin{figure}[ht]
\centering
\includegraphics[width=90mm , height=80mm]{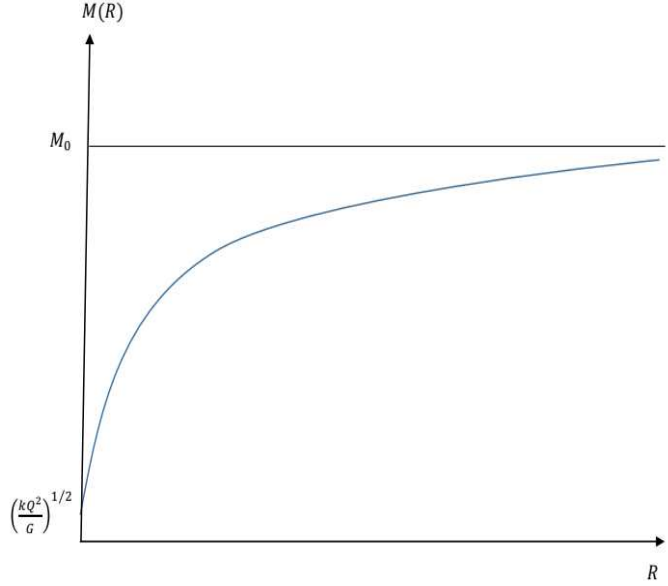}
\caption[Graph of the total mass vs R.]{Graph of the total mass $M$ as a function of $R$ . }
\label{fig:MvsR}
\end{figure}

where $a=\frac{3}{5}$. We note that $M$ is a function of $R$. From equation (\ref{eq:M}), we can  see that $lim_{R\rightarrow \infty}\left(M\left(R\right)\right)=M_{0}$ and  $lim_{R\rightarrow 0}\left(M\left(R\right)\right)=\left(\frac{kQ^{2}}{G}\right)^{\frac{1}{2}}$, or $\frac{GM}{c^{2}} \rightarrow \frac{Q }{c^{2}} \sqrt{\frac{G}{4\pi\epsilon_{0}}}$, as expected (see figure \ref{fig:MvsR}). The derivative of $M$ with respect to $R$ give us:

\begin{eqnarray}
\frac{dM}{dR}=\frac{a\left(GM^{2}-kQ^{2}\right)}{2aGMR+R^{2}c^{2}}
\end{eqnarray}
To have collapse, we need that $\frac{dM}{dR}>0$. Then collapse occurs only if: $GM^{2}>kQ^{2}$, or $\frac{GM}{c^{2}} > \frac{Q }{c^{2}} \sqrt{\frac{G}{4\pi\epsilon_{0}}}$, as we had expected.

\section{Equation of motion}
\label{equation of motion}

Lets consider a gravitational source (like a black hole) of mass $M$ and charge $Q$ and a massive particle that is moving around the source. The motion of such particle is governed by the Reissner-Nordstr\"om metric.  Because the orbital angular momentum must be constant,  the orbital motion will be performed in a single plane. Then, we can analyze the motion of the particle in the equatorial plane $\theta=\pi/2$:

\begin{equation}
(ds)^{2} = c^{2}(d\tau)^{2}=\Delta c^{2}(dt)^{2}-\Delta^{-1}(dr)^{2}-r^{2}(d\phi)^{2}.\label{eq:Seg2}
\end{equation}

The geodesic equation can be written in an alternative form using the Lagrangian 

\begin{eqnarray}
L\left(x^{\mu}, \frac{dx^{\mu}}{d\sigma}\right) = - g_{\alpha \beta}\left(x^{\mu}\right)
\frac{dx^{\alpha}}{d\sigma}\frac{dx^{\beta}}{d\sigma} -2\frac{q}{m} A_{\alpha}\frac{dx^{\alpha}}{d\sigma}
\label{eq:Lagrangian RN}
\end{eqnarray}

where $\sigma$ is a parameter of the trajectory of the particle, which is usually taken to be the proper time, $\tau$ for a massive particle. $A_{\alpha}$ is the potential four-vector, $q$ is the charge of the particle and $m$ is its rest-mass. Using the Euler-Lagrange equations:

\begin{eqnarray}
\frac{\partial L}{\partial x^{\mu}}-\frac{d}{d \sigma}\left(\frac{\partial L}{\partial \left(\frac{dx^{\mu}}{d \sigma}\right)}\right)=0
\end{eqnarray}

we get the geodesic equation for the particle:

\begin{eqnarray}
\frac{du_{\mu}}{d\sigma} = \frac{1}{2} \left(\partial_{\mu} g_{\alpha \beta}\right) u^{\alpha} u^{\beta} + \frac{q}{m} F_{\mu \alpha}u^{\alpha}  \label{eq:geodesic}
\end{eqnarray}
where $u_{\mu} = \frac{dx_{\mu}}{d \sigma}$, and $F_{\mu \alpha}=\partial_{\mu} A_{\alpha}-
\partial_{\alpha} A_{\mu}$ is an antisymmetric tensor. In the present case: 

\begin{equation}
F_{01}=-F_{10}= \frac{E^{1}}{c}=\frac{E\left(r\right)}{c},
\end{equation}

the other $F_{\mu \nu}=0$ (we are considering a static electric field $\vec{E} = E\left(r\right) \vec{e_{r}}$ and we take $\vec{B}=\vec{0}$).

For the coordinates $ct$ ($\mu = 0$) and  $\phi$ ($\mu =3$) the geodesic equation  (\ref{eq:geodesic}) gives us, respectively :

\begin{eqnarray}
\frac{d}{d \sigma} \left[ \Delta  c \frac{d \left(ct\right)}{d \sigma}+\frac{qQ}{4\pi \epsilon_{0} m  r}\right] = 0 \label{eq:energy}
\end{eqnarray}
and
\begin{eqnarray}
\frac{d}{d \sigma}\left[m r^{2} \frac{d \phi}{d \sigma} \right] = 0. \label{eq:angularmom}
\end{eqnarray}

Equations (\ref{eq:energy}) and (\ref{eq:momang}) with $\sigma = \tau$ imply that there are two constants of motion.

\begin{equation}
E^{*}=\Delta c^{2} \frac{dt}{d\tau} +\frac{qQ}{4\pi \epsilon_{0} m  r}  \label{eq:energy1}
\end{equation}
and

\begin{equation}
J=r^{2}\frac{d\phi}{d\tau}.\label{eq:momang}
\end{equation}

The first constant is the energy per unit mass, meanwhile the second is the angular momentum per unit mass.

Returning to equation (\ref{eq:Seg2}), using   $\frac{dr}{d\tau} = \frac{dr}{d\phi} \frac{d\phi}{d\tau}$ , it can be written as:

\begin{eqnarray}
c^{2}=\Delta c^{2} \left( \frac{dt}{d\tau}\right)^{2}  - \Delta^{-1} \left(\frac{dr}{d \phi} \right)^{2}\left(\frac{d\phi}{d\tau}\right)^{2} - r^{2} \left(\frac{d\phi}{d\tau}\right)^{2}.  \label{eq:RN1}
\end{eqnarray}

From (\ref{eq:energy1}) and ( \ref{eq:momang}) we can express $\frac{dt}{d\tau}$ and $\frac{d\phi}{d\tau}$ in terms of $E^{*}$ and $J$ to obtain:

\begin{eqnarray}
\left(\frac{dr}{d \phi} \right)^{2}=- \Delta\left( \frac{c^{2}r^{4}}{J^{2}}+r^{2}\right)+\frac{r^{4}}{J^{2}c^{2}}\left(E^{*}-\frac{qQ}{4\pi \epsilon_{0} m  r}\right)^{2} \label{eq:orbit1}
\end{eqnarray}

This equation can be turned into an equation for $U(\phi) = \frac{1}{r(\phi)}$, noting that 
\begin{equation}
\frac{dr}{d\phi} = - \frac{1}{U^{2}} \frac{dU}{d\phi}
\end{equation}

so we arrive at the following equation for $U(\phi)$:

\begin{eqnarray}
\left(\frac{dU}{d\phi}\right)^{2}= - \Delta \left(\frac{c^{2}}{J^{2}}+U^{2}\right)+ \frac{1}{J^{2}c^{2}}\left(E^{*}-\frac{qQU}{4\pi \epsilon_{0} m  }\right)^{2}
 \label{eq:orbit2}
\end{eqnarray}

By taking the derivative of equation (\ref{eq:orbit1}) with respect to $\phi$, we get the following differential equation for $U \left(\phi \right)$:

\[ [\frac{dU}{d\phi}][\frac{d^{2} U}{d \phi^{2}} + U\left(1-\left(\frac{q Q}{4 \pi \epsilon_{0} m c J} \right)^{2} + \frac{Q^{2}G}{4 \pi \epsilon_{0} c^{2} J^{2}}\right)
\]

\[-\frac{3GMU^{2}}{c^{2}}-\frac{GM}{J^{2}}+\frac{Q^{2}GU^{3}}{2\pi \epsilon_{0}c^{4}}+\frac{qQE^{*}}{4 \pi \epsilon_{0} m J^{2}c^{2}}]=0
\]
\\

This differential equation can be separated into two differential equations for $U(\phi)$. The first one  is the equation for a particle traveling inwards or outwards of the source of the gravitational field:

\begin{equation}
\frac{dU}{d\phi} = 0,
\label{eq:orbit3}
\end{equation}

the other differential equation, applicable for trajectories in which $U(\phi)$ is not constant with respect to $\phi$, is the following:

\begin{equation}
\frac{d^{2} U}{d \phi^{2}} + U\left(1-\left(\frac{q Q}{4 \pi \epsilon_{0} m c J} \right)^{2} + \frac{Q^{2}G}{4 \pi \epsilon_{0} c^{2} J^{2}}\right)=
\frac{3GMU^{2}}{c^{2}}+\frac{GM}{J^{2}}-\frac{Q^{2}GU^{3}}{2\pi \epsilon_{0}c^{4}}-\frac{qQE^{*}}{4 \pi \epsilon_{0} m J^{2}c^{2}}
\label{eq:orbit4}
\end{equation}

This is the equation for the trajectory of a massive particle that travels around the source  in the equatorial plane.

For a photon, the geodesic equation is \cite{MarinRodriguez}

\begin{eqnarray}
\frac{d u_{\mu}}{d\lambda}=\frac{1}{2} \partial_{\mu}\left(g_{\alpha \beta}\right)u^{\alpha}u^{\beta},
\label{eq:photongeodesic}
\end{eqnarray}
where $\lambda$ is an afin parameter. 

Consider a photon traveling in the equatorial plane ($\theta = \frac{\pi}{2}$) around the source. For the coordinates $ct$ ($\mu = 0$) and  $\phi$ ($\mu =3$) the geodesic equation  (\ref{eq:photongeodesic}) give us, respectively,  the following constants along the trajectory of the photon around the source:

\begin{eqnarray}
E^{**}=\Delta c^{2} \left(\frac{dt}{d\lambda}\right)
\label{eq:photonenergy}
\end{eqnarray}
\begin{eqnarray}
J=r^{2}\left(\frac{d \phi}{d \lambda}\right) \label{eq:photonangular},
\end{eqnarray}
where $E^{**}$ has units of energy per unit mass and $J$ of angular momentum per unit mass (when $\lambda$ has units of time). The invariant interval can be written as:

\begin{eqnarray}
0= \Delta c^{2} \left( \frac{dt}{d\lambda}\right)^{2}  - \Delta^{-1} \left(\frac{dr}{d \phi} \right)^{2}\left(\frac{d\phi}{d\lambda}\right)^{2} - r^{2} \left(\frac{d\phi}{d\lambda}\right)^{2}.  \label{eq:RN2}
\end{eqnarray}
Inserting the definitions of $E^{**}$ and $J$ in (\ref{eq:RN2}) , with the change of variable
$U(\phi) = \frac{1}{r(\phi)}$ , and following a similar procedure to the one used to deduce the orbit equation for a massive particle, we get:

\[ \left(\frac{dU}{d\phi}\right)\left(\frac{d^{2} U}{d \phi^{2}} + U
-\frac{3GMU^{2}}{c^{2}}+\frac{Q^{2}GU^{3}}{2\pi \epsilon_{0}c^{4}}\right)=0
\]
\\
This differential equation can be separated into two differential equations for $U\left(\phi\right)$. The first one is the equation for a photon that travels directly into or out from the source:
\begin{eqnarray}
\left(\frac{dU}{d\phi}\right)=0. \label{eq:photonorbit1}
\end{eqnarray}
The other differential equation for trajectories in which $U\left(\phi\right)$ is not constant with respect to $\phi$, is:
\begin{eqnarray}
\frac{d^{2} U}{d \phi^{2}} + U=
\frac{3GMU^{2}}{c^{2}}-\frac{Q^{2}GU^{3}}{2\pi \epsilon_{0}c^{4}} \label{eq:photonorbit2}
\end{eqnarray}
The last one is the equation for the trajectory of a photon that travels around the source in the equatorial plane.

\section{Photon Surfaces}
\label{photon surfaces}

Let's consider the differential equation for a photon traveling around a massive charged object like a charged black hole (equation \ref{eq:photonorbit2}). This equation has exact constant solutions, for the unstable circular orbits of a photon around the source. In fact, for circular orbits we have $\frac{dU}{d\phi}=0$ and $\frac{d^{2}U}{d\phi^{2}}=0$, and then, equation (\ref{eq:photonorbit2}) for $U\neq 0$ give us:

\begin{eqnarray}
r_{1c,2c}=\frac{3GM}{c^{2}}\left(\frac{Q^{2}}{9 \pi \epsilon_{0} GM^{2}}\right)\left(1\pm \left(1-\frac{2Q^{2}}{9 \pi \epsilon_{0} GM^{2}}\right)^{\frac{1}{2}}\right)^{-1},
\label{eq:photonsphere1}
\end{eqnarray}
where $r_{1c}$ and $r_{2c}$ are the radii of the so-called photon spheres \cite{Claudel}, being $r_{2c}$ the outermost photon sphere radius. Equation (\ref{eq:photonsphere1}), also can be written as: 

\begin{eqnarray}
r_{1c,2c}=\frac{r_{c}}{2}\left(1\mp \left(1-8\frac{r_{Q}^{2}}{r_{c}^{2}}\right)^{\frac{1}{2}}\right),
\label{eq:photonsphere2}
\end{eqnarray}
where $r_{c}=\frac{3GM}{c^{2}}$ and $r_{Q}^{2}=\frac{Q^{2}G}{4\pi\epsilon_{0}c^{4}}$.

It can easily be shown that:

\begin{equation}
r_{1c}<r_{2c}<r_{c}=\frac{3GM}{c^{2}},
\label{eq:photonineq}
\end{equation}

where $r_{c}$ is the radius of the photon sphere for a photon traveling around a compact object (a black hole) in a space-time described by the Schwarzschild metric. In terms of the Schwarzschild radius, we can write  $r_{c}=\frac{3}{2} r_{s}$.

In terms of $r_{c}$ the orbit equation (\ref{eq:photonorbit2}) can be written as:

\begin{eqnarray}
\frac{d^{2} U}{d \phi^{2}} + U=r_{c}\left(U^{2}-\frac{Q^{2}U^{3}}{6 \pi \epsilon_{0} M c^{2}}\right).
\label{eq:orbitlightdev}
\end{eqnarray}

Consider the initial conditions shown in Figure \ref{fig:defl01}. The smallest value of the $r-$coordinate in the trajectory, $r=b$, is taken such that the photon escapes the black hole, $b>r_{2c}$. However because the difference between $r_{2c}$ and $r_{c}$ is very small, we will rewrite Equation (\ref{eq:orbitlightdev}) in terms of $\epsilon = \frac{r_c}{b} < 1$, which we will use as a non-dimensional small number for our following perturbative expansions. Note that by multiplying both sides of Equation (\ref{eq:orbitlightdev}) by $b$, and defining the non-dimensional trajectory parameter

\begin{figure}[ht]
\centering
\includegraphics[width=80mm , height=80mm]{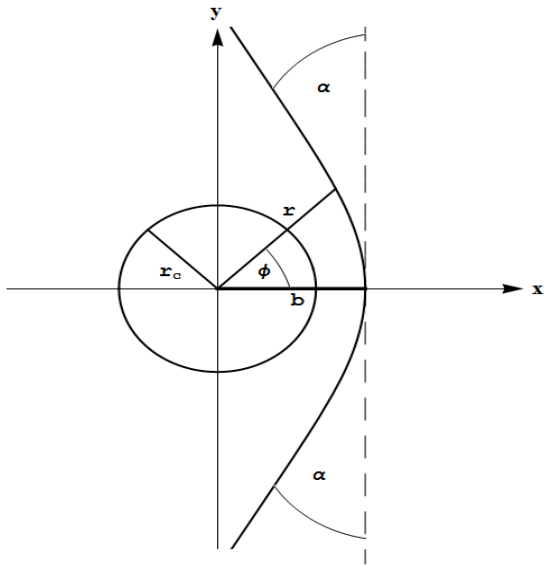}
\caption[Trajectory of a photon outside the photon sphere.]{Trajectory of a photon outside the photon sphere. The initial conditions are taken such that $r|_{\phi=0}=b$, and is called the impact parameter of the trajectory - the closest distance from the trajectory to the center of the black hole. We thus have, $\frac{dr}{d\phi}|_{\phi=0}=0$ and $\frac{dU}{d\phi}|_{\phi=0}=0 $. The photon experiences a total angular deflection of $2\alpha$. }
\label{fig:def01}
\end{figure}

\begin{equation}
V (\phi) = \frac{b}{r (\phi)}  
\label{eq:V}
\end{equation} 

Equation (\ref{eq:orbitlightdev}), with the inclusion of the term $\epsilon = \frac{r_{c}}{b}$, then becomes a differential equation  in $V(\phi)$:

\begin{equation}
\label{eq:deflection-1}
\frac{d^2 V (\phi)}{d\phi^2} +V(\phi) = \epsilon \left(V^2 (\phi)-\frac{\delta}{b}V^3 (\phi)\right)
\end{equation} 

where $0 < \epsilon = \frac{r_{c}}{b} < 1$, $\delta\equiv \frac{Q^{2}}{6\pi \epsilon_{0} M c^{2}} $ and with initial conditions given by

\begin{equation}
\label{eq:init-conditions-v}
V(\phi=0)=1\,;\,\frac{dV}{d\phi}(\phi=0)=0  
\end{equation}

Under these conditions, $V(\phi)$ is bounded such that

\begin{equation}
|V(\phi)|  \leq  1.
\end{equation}

\section{First-order solution for $V(\phi)$}
\label{First-order}

To obtain a solution of Equation (\ref{eq:deflection-1}) we can  write  $V(\phi)$ as a power series in $\epsilon$:

\begin{equation}
\label{eq:deflection-2}
V(\phi;\epsilon)= V_0 (\phi) + \epsilon V_1(\phi) +\epsilon^2 V_2(\phi) + ...
\end{equation}

Plugging the expansion (\ref{eq:deflection-2}) into Equation (\ref{eq:deflection-1}) results in the following:

\begin{eqnarray}
\label{eq:deflection-3}
\left(\frac{d^2 V_0}{d\phi^2} + \epsilon \frac{d^2 V_1}{d\phi^2} +  \epsilon^2 \frac{d^2 V_2}{d\phi^2} + ...\right) +(V_0 + \epsilon V_1 + \epsilon^2 V_2+...)  \nonumber
\end{eqnarray}

\begin{eqnarray} 
= \epsilon \left(V_0 + \epsilon V_1 + \epsilon^2 V_2+...\right)^2 
- \epsilon \frac{\delta}{b}  \left(V_0 + \epsilon V_1 + \epsilon^2 V_2+...\right)^3 . 
\end{eqnarray}

We can group the powers of $\epsilon$ in Equation (\ref{eq:deflection-3}):

\begin{eqnarray}
  \epsilon^0 :&   \frac{d^2 V_0}{d\phi^2} + V_0 = 0      \label{eq:def6a}  
\end{eqnarray}

\begin{eqnarray}
  \epsilon^1 :&   \frac{d^2 V_1}{d\phi^2} + V_1 = {V_0}^2\left(1-\frac{\delta}{b}V_0\right)       \label{eq:def6b}     \end{eqnarray}
  
\begin{eqnarray} 
  \epsilon^2 :&   \frac{d^2 V_2}{d\phi^2} + V_2 = V_{0}\left(2 V_{1} - 3 \frac{\delta}{b}V_{0}V_{1}       \right) \label{eq:def6c} \end{eqnarray}

  \begin{eqnarray}
  &\vdots   \nonumber
\end{eqnarray}

 The initial conditions of $V(\phi)$, applied to the asymptotic expansion in Equation (\ref{eq:deflection-2}), imply (by grouping powers of $\epsilon$):

\begin{eqnarray}
  \epsilon^0 :&   V_0(0)=1 \,;\, \frac{dV_0}{d\phi}(0)=0      \label{eq:inita} 
\end{eqnarray}
  
\begin{eqnarray}
  \epsilon^k :&   V_k(0)=0 \,;\, \frac{dV_k}{d\phi}(0)=0 \,;\, k\geq1 . \label{eq:initb} 
\end{eqnarray}

From these differential equations and initial conditions, we can readily obtain $V_0$ and $V_1$ iteratively: 

\begin{equation}
V_0(\phi)= \cos (\phi)
\label{eq:sol1}
\end{equation}

\begin{eqnarray}
\label{eq:sol2}
V_1 (\phi)= \frac{2}{3}+ \left(-\frac{1}{3}+\frac{\delta}{4b}\right) \cos (\phi)-\frac{1}{3} \cos^2(\phi)
-\frac{\delta}{4b} \cos^{5}(\phi) \nonumber 
\end{eqnarray}
\begin{eqnarray}
-\frac{\delta}{2b}\left(\frac{3\phi}{4}+\frac{5}{4} \sin(\phi)\cos(\phi)-\frac{1}{2}\sin^{3}(\phi)\cos(\phi)\right)\sin(\phi)
\end{eqnarray}

Thus, we obtain an equation for $V(\phi)$, per Equation (\ref{eq:deflection-2}):

\begin{eqnarray}
\label{eq:soltotal}
V(\phi) = cos (\phi) + \epsilon (\frac{2}{3}+ \left(-\frac{1}{3}+\frac{\delta}{4b}\right) \cos (\phi)-\frac{1}{3} \cos^2(\phi)-\frac{\delta}{4b} \cos^{5}(\phi) \nonumber 
\end{eqnarray}
\begin{eqnarray}
-\frac{\delta}{2b}\left(\frac{3\phi}{4}+\frac{5}{4} \sin(\phi)\cos(\phi)-\frac{1}{2}\sin^{3}(\phi)\cos(\phi)\right)\sin(\phi))+O(\epsilon^2).
\end{eqnarray}

According to the coordinate system shown in Figure 1, the photon goes through a total angular deflection of $2\alpha$. This corresponds to setting $V(\phi)=0$ for both $\phi=\pi/2+\alpha$ and $\phi=-\pi/2-\alpha$. From both of these conditions considering that $\alpha$ is very small, to first order in $\epsilon$ we get: 
\begin{eqnarray}
\alpha = \frac{\epsilon\left(\frac{2}{3}-\frac{3\pi\delta}{16b}\right)}{\left(1 - \frac{\epsilon}{3}+\frac{\epsilon \delta}{4b}\right)} . \label{eq:alpha1}
\end{eqnarray}
The total deviation of the photon is then
\begin{equation}
\Omega = 2\alpha  \approx  \epsilon\left(\frac{4}{3}-\frac{3\pi \delta}{8b}\right) = \frac{r_{c}}{b}\left(\frac{4}{3}-\frac{3\pi \delta}{8b}\right)= \frac{4GM}{bc^{2}}-\frac{3Q^{2}G}{16b^{2}\epsilon_{0}c^{4}} .
\label{eq:alpha2}
\end{equation}

Note that $\Omega <  \frac{4GM}{bc^{2}}$. Then,  the angle of deviation is smaller than the one corresponding  to a black hole without electric charge. This result agrees with the calculated by 
Shchigolev-Bezbatko  \cite{Shchigolev} and Briet-Hobill \cite{Briet}.

\section{Towards a second-order solution for $\Omega(\epsilon)$}
\label{Towards a second-order}

We will now see how to obtain a second order solution for $\Omega$. 

The term in Equation (\ref{eq:soltotal}) that goes as $\phi sin(\phi)$ grows without bound. Then, if we naively include these kind of terms (called secular terms \cite{Bush}), our solution is no longer bounded. Thus, we have to eliminate any and all secular term that arises to arrive at a well-behaved solution for $V(\phi)$. \\
One method to do this, due to Lindstedt and Poincar\'e, is by solving the differential equation in the following \textit{strained coordinate} \cite{Bush}:

\begin{equation}
\label{eq:strained}
\tilde{\phi}= \phi \left( 1 + \omega_1 \epsilon + \omega_2 \epsilon^2 + \ldots \right) .
\end{equation}

Where the $\omega_k$ are constants to be determined. In terms of this new strained coordinate $\tilde{\phi}$, Equation (\ref{eq:deflection-1}) becomes

\begin{eqnarray}
\label{eq:dif-V-tilde}
\left(1 + \omega_1 \epsilon + \omega_2 \epsilon^2 + \ldots \right)^2 \frac{d^2V}{d\tilde{\phi}^2} + V(\tilde{\phi}) = \epsilon \left( V^2(\tilde{\phi})-\frac{\delta}{b}V^3(\tilde{\phi})\right) .
\end{eqnarray}

Assuming an asymptotic expansion on $V(\tilde{\phi})$:

\begin{equation}
\label{eq:expansion-V-tilde}
V(\tilde{\phi};\epsilon)= V_0 (\tilde{\phi}) + \epsilon V_1(\tilde{\phi}) +\epsilon^2 V_2(\tilde{\phi}) + ...
\end{equation}  \\

Plugging  the expansion(\ref{eq:expansion-V-tilde}) in Equation (\ref{eq:dif-V-tilde}), we obtain:

\begin{eqnarray}
\label{eq:V-tilde-expansion}
\left(1 + \omega_1 \epsilon + \omega_2 \epsilon^2 + \ldots \right)^2 \left(\frac{d^2 V_0}{d\tilde{\phi}^2}   + \epsilon \frac{d^2 V_1}{d\tilde{\phi}^2} +  \epsilon^2 \frac{d^2 V_2}{d\tilde{\phi}^2} + ...\right)  + \nonumber \\
+(V_0 + \epsilon V_1 + \epsilon^2 V_2+...)  
  = \epsilon \left(V_0 + \epsilon V_1 + \epsilon^2 V_2+...\right)^2  \nonumber \\
  -\epsilon \frac{\delta}{b}\left((V_0 + \epsilon V_1 + \epsilon^2 V_2+...\right)^3
\end{eqnarray}

   We can group the powers of $\epsilon$ in Equation (\ref{eq:V-tilde-expansion}):

\begin{eqnarray}
  \epsilon^0 :&   \frac{d^2 V_0}{d\tilde{\phi}^2} + V_0 = 0      \label{eq:Vdif0} \end{eqnarray}
  
 \begin{eqnarray}
  \epsilon^1 :&   \frac{d^2 V_1}{d\tilde{\phi}^2} + V_1 = {V_0}^2 - 2 \omega_1  \frac{d^2 V_0}{d\tilde{\phi}^2}-\frac{\delta}{b}V_{0}^{3}       \label{eq:Vdif1}
  \end{eqnarray}
  
  \begin{eqnarray}
  \epsilon^2 :&   \frac{d^2 V_2}{d\tilde{\phi}^2} + V_2 = 2 V_0 V_1 - ({\omega_1}^2+2\omega_2 ) \frac{d^2 V_0}{d\tilde{\phi}^2} - 2 \omega_1 \frac{d^2 V_1}{d \tilde{\phi}^2}-3\frac{\delta}{b}V_{0}^{2} V_{1}        \label{eq:Vdif2} 
  \end{eqnarray}
  
  \begin{eqnarray}
  \epsilon^3 :&   \frac{d^2 V_3}{d\tilde{\phi}^2} + V_3 = {V_1}^2 +  2 V_0 V_2 - (2 \omega_1 \omega_2 + 2 \omega_3) \frac{d^2 V_0}{d\tilde{\phi}^2}  \nonumber  
 \end{eqnarray}
 \begin{eqnarray}
  - ( {\omega_1}^2+2\omega_2)\frac{d^2 V_1}{d\tilde{\phi}^2} - 2 \omega_1 \frac{d^2 V_2}{d\tilde{\phi}^2}-\frac{3\delta}{b}V_{0}\left(V_{0}V_{2}+V_{1}^{2}\right) .      \label{eq:Vdif3} 
  \end{eqnarray} 
  \begin{eqnarray}
  &\vdots   \nonumber
  \end{eqnarray}
  
     With some care due to the definitions of the scaled variable and its derivative, we arrive at initial conditions for the $V_k(\tilde{\phi})$ from the initial conditions of $V(\phi)$:

\begin{eqnarray}
  \epsilon^0 :&   V_0( \tilde{\phi}=0)=1 \,;\, \frac{dV_0}{d\tilde{\phi}}(0)=0      \label{eq:init-tilde-a}
 \end{eqnarray}
 
 \begin{eqnarray}
  \epsilon^k :&   V_k(\tilde{\phi}=0)=0 \,;\, \frac{dV_k}{d\tilde{\phi}}(0)=0 \,;\, k\geq1 \label{eq:init-tilde-b}\
\end{eqnarray}

Solving the differential equation (\ref{eq:Vdif0}) with initial conditions (\ref{eq:init-tilde-a}), we arrive at the zeroth-order  contribution to $V(\tilde{\phi})$:

\begin{equation}
\label{eq:othorder}
V_0 (\tilde{\phi}) = cos(\tilde{\phi})
\end{equation}

Introducing the value of $V_{0}$ given by (\ref{eq:othorder}) in equation (\ref{eq:Vdif1}), we get the differential equation:

\begin{eqnarray} 
\frac{d^2 V_1}{d\tilde{\phi}^2} + V_1 =\cos^{2}(\tilde{\phi})\left(1-\frac{\delta}{b}\cos(\tilde{\phi})\right)+2\omega_{1} \cos(\tilde{\phi})
\label{eq:firstorder}
\end{eqnarray}

The general solution of the equation (\ref{eq:firstorder}) is:

\begin{eqnarray}
V_{1}(\tilde{\phi})=C_{1}\cos(\tilde{\phi})+C_{2}\sin(\tilde{\phi})+\frac{1}{3}\cos^{4}(\tilde{\phi})
-\frac{\delta}{4b}\cos^{5}(\tilde{\phi})+\omega_{1}\cos^{3}(\tilde{\phi})  \nonumber
\end{eqnarray}
\begin{eqnarray}
\quad +\sin^{2}(\tilde{\phi})-\frac{1}{3}\sin^{4}(\tilde{\phi})-\frac{3\delta}{8b}\tilde{\phi}\sin(\tilde{\phi})
-\frac{5\delta}{8b}\sin^{2}(\tilde{\phi})\cos(\tilde{\phi}) \nonumber
\end{eqnarray}
\begin{eqnarray}
\quad +\frac{\delta}{4b}\sin^{4}(\tilde{\phi})cos(\tilde{\phi})+\omega_{1}\tilde{\phi}\sin(\tilde{\phi})
+\omega_{1}\sin^{2}(\tilde{\phi})\cos(\tilde{\phi})
\end{eqnarray}
The conditions (\ref{eq:init-tilde-b}) for $k=1$ give us:
\begin{eqnarray}
C_{1}=-\frac{1}{3}+\frac{\delta}{4b}-\omega_{1}
\end{eqnarray}
and
\begin{eqnarray}
C_{2}=0
\end{eqnarray}
and then:
\begin{eqnarray}
V_{1}(\tilde{\phi})=\left(-\frac{1}{3}+\frac{\delta}{4b}-\omega_{1}\right) \cos(\tilde{\phi})+\frac{1}{3}\cos^{4}(\tilde{\phi})
-\frac{\delta}{4b}\cos^{5}(\tilde{\phi})+\omega_{1}\cos^{3}(\tilde{\phi})  \nonumber
\end{eqnarray}
\begin{eqnarray}
\quad +\sin^{2}(\tilde{\phi})-\frac{1}{3}\sin^{4}(\tilde{\phi})+\left(\omega_{1}-\frac{3\delta}{8b}\right)\tilde{\phi}\sin(\tilde{\phi})
-\frac{5\delta}{8b}\sin^{2}(\tilde{\phi})\cos(\tilde{\phi}) \nonumber
\end{eqnarray}
\begin{eqnarray}
\quad +\frac{\delta}{4b}\sin^{4}(\tilde{\phi})cos(\tilde{\phi})
+\omega_{1}\sin^{2}(\tilde{\phi})\cos(\tilde{\phi}).
\label{eq:secondorder1}
\end{eqnarray}
As we can check, a secular term has appeared for $V_{1}(\tilde{\phi})$ (a term that goes as $\tilde{\phi}\sin(\tilde{\phi})$). Then, equation (\ref{eq:secondorder1}) grows without limit.  However, we can use the freedom in the definition of $\omega_{1}$ to eliminate this secular term by setting:
\begin{eqnarray}
\omega_{1}=\frac{3\delta}{8b},
\end{eqnarray}
then, the final expression for $V_{1}(\tilde{\phi})$ is:
\begin{eqnarray}
V_{1}(\tilde{\phi})=\frac{2}{3}-\left(\frac{1}{3}+\frac{\delta}{8b}\right)\cos(\tilde{\phi})-\frac{1}{3}\cos^{2}(\tilde{\phi})+\frac{\delta}{8b}\cos^{3}(\tilde{\phi})  \label{eq:V1phi}
\end{eqnarray}
Thus, to first order $V_{1}\left(\tilde{\phi};\epsilon\right)$ is:
\begin{eqnarray}
\label{eq:Vtildefirstorder}
V_{1}(\tilde{\phi};\epsilon)=\cos(\tilde{\phi})+\epsilon\left(\frac{2}{3}-\left(\frac{1}{3}+\frac{\delta}{8b}\right)\cos(\tilde{\phi})-\frac{1}{3}\cos^{2}(\tilde{\phi})+\frac{\delta}{8b}\cos^{3}(\tilde{\phi})\right)
\end{eqnarray}
We set up $\tilde{\phi}=\frac{\pi}{2}+\tilde{\alpha}$ in equation (\ref{eq:Vtildefirstorder}), such that $V\left(\frac{\pi}{2}+\tilde{\alpha}\right)=0$. We obtain the equation:
\begin{eqnarray}
\label{eq:sinalpha}
-\sin(\tilde{\alpha})+\epsilon\left(\frac{2}{3}+\left(\frac{1}{3}+\frac{\delta}{8b}\right)\sin(\tilde{\alpha})
-\frac{1}{3}\sin^{2}(\tilde{\alpha})-\frac{\delta}{8b}\sin^{3}(\tilde{\alpha})\right)+O(\epsilon^{3})=0.
\end{eqnarray}
To solve this equation let's assume that $\sin(\tilde{\alpha})$ has the following expansion in $\epsilon$, with a leading order term of order $\epsilon^{1}$:

\begin{eqnarray}
\label{eq:expansion1}
sin(\tilde{\alpha}) = \epsilon \chi_1 + \epsilon^2 \chi_2 + \epsilon^3 \chi_3 + \ldots
\end{eqnarray}
where the $\chi_k$ are constants to be determined. Inserting this new expansion into Equation (\ref{eq:sinalpha}) leads to the following algebraic equation to first order:
\begin{eqnarray}
0=-\epsilon \left(-\chi_{1}+\frac{2}{3}\right),
\end{eqnarray}
from which equating to zero the terms with $\epsilon^1$  we arrive at:
\begin{equation}
\chi_{1}=\frac{2}{3}
\end{equation}
Thus, to first order,  $sin(\tilde{\alpha})$ is given by:

\begin{equation}
sin(\tilde{\alpha}) = \frac{2}{3} \epsilon 
\end{equation}
To obtain $\tilde{\alpha}$, we employ the Taylor series of $\arcsin(x)$ around $x=0$:

\begin{equation}
\arcsin(x) = x +\frac{x^3}{6} + O(x^5)
\end{equation}
and obtain
\begin{equation}
\label{eq:alphatecho}
\tilde{\alpha}=\frac{2}{3}\epsilon
\end{equation}
From the definition of the strained coordinate $\tilde{\phi}$ in (\ref{eq:strained}), it is clear that:

\begin{equation}
\frac{\pi}{2} + \alpha=\frac{\frac{\pi}{2} + \tilde{\alpha}}{1+\omega_1 \epsilon + \omega_2 \epsilon^2 + O(\epsilon^3)} , 
\end{equation}
and to the first order we have:
\begin{eqnarray}
\frac{\pi}{2} + \alpha=\frac{\frac{\pi}{2} + \tilde{\alpha}}{1+ \frac{3\delta}{8b}\epsilon} \approx \frac{\pi}{2} + \tilde{\alpha}-\frac{3\delta\pi}{16b}\epsilon , 
\end{eqnarray}
and replacing the value of  $\tilde{\alpha}$ given by ( \ref{eq:alphatecho}), we get:
\begin{eqnarray}
\alpha=\left(\frac{2}{3}-\frac{3\delta\pi}{16b}\right)\epsilon.
\end{eqnarray}
So, the total deflection angle is to first order:
\begin{eqnarray}
\Omega = 2 \alpha = \left(\frac{4}{3}-\frac{3\delta\pi}{8b}\right)\epsilon
\end{eqnarray}
Introducing the values of $\epsilon$ and $\delta$, we finally get:
\begin{eqnarray}
\Omega =  \frac{4GM}{bc^{2}}-\frac{3Q^{2}G}{16b^{2}\epsilon_{0}c^{4}} ,
\label{eq:alpha1order}
\end{eqnarray}
which matches the value obtained in equation (\ref{eq:alpha2}).

\section{Second-order solution for $\Omega(\epsilon)$}
\label{Second-order}

Let´s return to equation (\ref{eq:Vdif2}), but now let´s rewrite it in the form:
  \begin{eqnarray}
  \epsilon^2 :&   \frac{d^2 V_2}{d\tilde{\phi}^2} + V_2 =V^{*}(\tilde{\phi}) ,   \label{eq:V2} 
  \end{eqnarray}
  where
\begin{eqnarray}
V^{*}(\tilde{\phi})=2 V_0 V_1 - ({\omega_1}^2+2\omega_2 ) \frac{d^2 V_0}{d\tilde{\phi}^2} - 2 \omega_1 \frac{d^2 V_1}{d \tilde{\phi}^2}-3\frac{\delta}{b}V_{0}^{2} V_{1}   \label{eq:V2*}
\end{eqnarray}
Introducing the expressions of $V_0(\tilde{\phi})$ and  $V_1(\tilde{\phi})$ given in (\ref{eq:othorder}) and (\ref{eq:V1phi}), respectively, we have:

\begin{eqnarray}
V^{*}(\tilde{\phi})=\left(\frac{4}{3}+2\omega_{2}-\frac{\delta}{4b}-\frac{33}{64}\left(\frac{\delta}{b}\right)^{2}\right)\cos(\tilde{\phi})-\left(\frac{2}{3}+\frac{13\delta}{4b}\right)\cos^{2}(\tilde{\phi})
\nonumber
\label{eq:V^*}
\end{eqnarray}

\begin{eqnarray}
+\left(-\frac{2}{3}+\frac{\delta}{b}+\frac{39}{32}\left(\frac{\delta}{b}\right)^{2}\right)\cos^{3}(\tilde{\phi})+\frac{5\delta}{4b}\cos^{4}(\tilde{\phi})-\frac{3}{8}\left(\frac{\delta}{b}\right)^{2}\cos^{5}(\tilde{\phi})+\frac{\delta}{2b} \label{eq:V*phitecho} 
\end{eqnarray}
The homogeneus equation corresponding to (\ref{eq:V2} )is:
\begin{equation}
\frac{d^2 V_2}{d\tilde{\phi}^2} + V_2 =0 ,    
\label{eq:homogV2}
\end{equation}
with general solution:
\begin{equation}
V_{2g}(\tilde{\phi})=c_{1}\cos(\tilde{\phi})+c_{2}\sin(\tilde{\phi})
\label{eq:V2g}
\end{equation}

Let´s look for a particular solution of the inhomogeneous equation (\ref{eq:V2}) of the form:

\begin{eqnarray}
V_{2p}(\tilde{\phi})=d_{1}(\tilde{\phi})\cos(\tilde{\phi})+d_{2}(\tilde{\phi})\sin(\tilde{\phi}),
\label{eq:V2p}
\end{eqnarray}
 satisfying:
 \begin{eqnarray}
\cos(\tilde{\phi})d_{1}^{'}+\sin(\tilde{\phi})d_{2}^{'}=0,
 \label{eq:V2pp}
 \end{eqnarray}
 where $d_{1}^{'}=\frac{d(d_{1}(\tilde{\phi}))}{d\tilde{\phi}}$ and $d_{2}^{'}=\frac{d(d_{2}(\tilde{\phi}))}{d\tilde{\phi}}$.
Introducing  $V_{2p}(\tilde{\phi})$ in equation (\ref{eq:V2}) we get:

\begin{eqnarray}
-\sin(\tilde{\phi})d_{1}^{'}+\cos(\tilde{\phi})d_{2}^{'}=V^{*}(\tilde{\phi}).
\label{eq:V2ppp}
\end{eqnarray}
\ref{eq:V2pp} and  \ref{eq:V2ppp} can be written in  matrix form as:

\begin{eqnarray}
\left(\begin{array}{cc}  \cos(\tilde{\phi}) & \sin(\tilde{\phi})\\ -\sin(\tilde{\phi}) & \cos(\tilde{\phi}) \end{array}\right)
\left(\begin{array}{c} d_{1}^{'} \\ d_{2}^{'} \end{array}\right) =
\left(\begin{array}{c} 0 \\ V^{*}(\tilde{\phi}) \end{array}\right) .
\end{eqnarray}

The solutions to this system of equations are:
\begin{equation}
d_{1}^{'}=-\sin(\tilde{\phi})V^{*}(\tilde{\phi}) ,
\label{eq:sol1}
\end{equation}
and
\begin{equation}
d_{2}^{'}=\cos(\tilde{\phi})V^{*}(\tilde{\phi}) .
\label{eq:sol2}
\end{equation}
Replacing the value of $V^{*}(\tilde{\phi})$  given by (\ref{eq:V^*}) and performing the integrals, we get:

\begin{eqnarray}
d_{1}(\tilde{\phi})= \frac{1}{2}\left(\frac{4}{3}+2\omega_{2}-\frac{\delta}{4b}-\frac{33}{64}\left(\frac{\delta}{b}\right)^{2}\right)\cos^{2}(\tilde{\phi})-\frac{1}{3}\left(\frac{2}{3}+\frac{13\delta}{4b}\right)\cos^{3}(\tilde{\phi})
\nonumber
\label{eq:d1}
\end{eqnarray}
\begin{eqnarray}
+\frac{1}{4}\left(-\frac{2}{3}+\frac{\delta}{b}+\frac{39}{32}\left(\frac{\delta}{b}\right)^{2}\right)\cos^{4}(\tilde{\phi})+\frac{\delta}{4b}\cos^{5}(\tilde{\phi})
\nonumber
\end{eqnarray}
\begin{eqnarray}
-\frac{1}{16}\left(\frac{\delta}{b}\right)^{2}\cos^{6}(\tilde{\phi})+\frac{\delta}{2b}\cos(\tilde{\phi}) ,
\end{eqnarray}
and
\begin{eqnarray}
d_{2}(\tilde{\phi})= \frac{\delta}{2b}\sin(\tilde{\phi})+\frac{1}{2}\left(\tilde{\phi}+\frac{1}{2}\sin(2\tilde{\phi})\right)\left(\frac{4}{3}+2\omega_{2}-\frac{\delta}{4b}-\frac{33}{64}\left(\frac{\delta}{b}\right)^{2}\right) 
\nonumber
\label{eq:d2}
\end{eqnarray}
\begin{eqnarray}
 -\left(\sin\tilde{\phi}-\frac{1}{3}\sin^{3}\tilde{\phi}\right)\left(\frac{2}{3}+\frac{13\delta}{4b}\right)+\left(\frac{3}{8}\tilde{\phi}+\frac{1}{4}\sin(2\tilde{\phi})+\frac{1}{32}\sin(4\tilde{\phi})\right)
 \nonumber
 \end{eqnarray}
 \begin{eqnarray}
\times \left(-\frac{2}{3}+\frac{\delta}{b}+\frac{39}{32}\left(\frac{\delta}{b}\right)^{2}\right)+\left(\sin(\tilde{\phi})-\frac{2}{3}\sin^{3}(\tilde{\phi})+\frac{1}{5}\sin^{5}(\tilde{\phi})\right)\left(\frac{5\delta}{4b}\right)
\nonumber
\end{eqnarray}
\begin{eqnarray}
-\frac{3}{64}\left(\frac{5}{2}\tilde{\phi}+2\sin(2\tilde{\phi})+\frac{3}{8}\sin(4\tilde{\phi})-\frac{1}{6}
\sin^{3}(2\tilde{\phi})\right)\left(\frac{\delta}{b}\right)^{2} .
\end{eqnarray}
Then, the general solution of equation (\ref{eq:V2}) is:

\begin{eqnarray}
V_{2}(\tilde{\phi})=c_{1}\cos(\tilde{\phi})+c_{2}\sin(\tilde{\phi})+d_{1}(\tilde{\phi})\cos(\tilde{\phi})+d_{2}(\tilde{\phi})\sin(\tilde{\phi}) ,
\label{eq:V2general}
\end{eqnarray}
where $d_{1}(\tilde{\phi})$ and $d_{2}(\tilde{\phi})$ are given by (\ref{eq:d1}) and (\ref{eq:d2}), respectively. $c_{1}$ and $c_{2}$ are constants to be determined by the conditions: $V_{2}(\tilde{\phi}=0)=0$ and $\frac{dV_{2}}{d\tilde{\phi}}(\tilde{\phi}=0)=0$. The result is:
\begin{eqnarray}
c_{1}=-\frac{5}{18}-\omega_{2}+\frac{5}{24}\left(\frac{\delta}{b}\right)+\frac{1}{64}\left(\frac{\delta}{b}\right)^{2}
\label{eq:d1constant}
\end{eqnarray}
\begin{eqnarray}
c_{2}=0.
\label{eq:d2constant}
\end{eqnarray}
Introducing the values of $c_{1}$ and $c_{2}$ in (\ref{eq:V2general}) we have:
\begin{eqnarray}
V_{2}(\tilde{\phi})=-\frac{4}{9}+\frac{5}{36}\cos(\tilde{\phi})+\frac{2}{9}\cos^{2}(\tilde{\phi})
+\frac{1}{12}\cos^{3}(\tilde{\phi})+\tilde{\phi}\sin(\tilde{\phi})\left(\omega_{2}+\frac{5}{12}\right)
\label{eq:V2casifinal} 
\nonumber
\end{eqnarray}
\begin{eqnarray}
+\left(\frac{\delta}{b}\right)\left(-1+\frac{11}{24}\cos(\tilde{\phi})+\frac{3}{4}\cos^{2}(\tilde{\phi})-\frac{1}{8}\cos^{3}(\tilde{\phi})-\frac{1}{12}\cos^{4}(\tilde{\phi})+\frac{1}{4}\tilde{\phi}\sin(\tilde{\phi})\right)
\nonumber
\end{eqnarray}
\begin{eqnarray}
+\left(\frac{\delta}{b}\right)^{2}\left(\frac{25}{256}\cos(\tilde{\phi})-\frac{29}{256}\cos^{3}(\tilde{\phi})+\frac{1}{64}\cos^{5}(\tilde{\phi})+\frac{21}{256}\tilde{\phi}\sin(\tilde{\phi})\right),
\end{eqnarray}
or
\begin{eqnarray}
V_{2}(\tilde{\phi})=\tilde{\phi}\sin(\tilde{\phi})\left(\omega_{2}+\frac{5}{12}+\frac{1}{4}\left(\frac{\delta}{b}\right)+\frac{21}{256}\left(\frac{\delta}{b}\right)^{2}\right)+ ...........
\label{eq:secularterm}
\end{eqnarray}
To eliminate the secular term in the last equation (the term containing $\tilde{\phi}\sin(\tilde{\phi})$), we set:
\begin{eqnarray}
\omega_{2}=-\left(\frac{5}{12}+\frac{1}{4}\left(\frac{\delta}{b}\right)+\frac{21}{256}\left(\frac{\delta}{b}\right)^{2}\right),
\end{eqnarray}
getting
\begin{eqnarray}
V_{2}(\tilde{\phi})=-\frac{4}{9}+\frac{5}{36}\cos(\tilde{\phi})+\frac{2}{9}\cos^{2}(\tilde{\phi})
+\frac{1}{12}\cos^{3}(\tilde{\phi})
\label{eq:V2final} 
\nonumber
\end{eqnarray}
\begin{eqnarray}
+\left(\frac{\delta}{b}\right)\left(-1+\frac{11}{24}\cos(\tilde{\phi})+\frac{3}{4}\cos^{2}(\tilde{\phi})-\frac{1}{8}\cos^{3}(\tilde{\phi})-\frac{1}{12}\cos^{4}(\tilde{\phi})\right)
\nonumber
\end{eqnarray}
\begin{eqnarray}
+\left(\frac{\delta}{b}\right)^{2}\left(\frac{25}{256}\cos(\tilde{\phi})-\frac{29}{256}\cos^{3}(\tilde{\phi})+\frac{1}{64}\cos^{5}(\tilde{\phi})\right).
\end{eqnarray}
Then, to second order we can write:
\begin{eqnarray}
V_{2}(\tilde{\phi};\epsilon)=\cos(\tilde{\phi})+\epsilon\left(\frac{2}{3}-\frac{1}{3}\cos(\tilde{\phi})-\frac{1}{3}\cos^{2}(\tilde{\phi})+\frac{\delta}{8b}\left(-\cos(\tilde{\phi})+\cos^{3}(\tilde{\phi})\right)\right)
\label{eq:Vfinal}
\nonumber
\end{eqnarray}
\begin{eqnarray}
+\epsilon^{2}(-\frac{4}{9}+\frac{5}{36}\cos(\tilde{\phi})+\frac{2}{9}\cos^{2}(\tilde{\phi})
+\frac{1}{12}\cos^{3}(\tilde{\phi})
\nonumber
\end{eqnarray}
\begin{eqnarray}
+\left(\frac{\delta}{b}\right)\left(-1+\frac{11}{24}\cos(\tilde{\phi})+\frac{3}{4}\cos^{2}(\tilde{\phi})-\frac{1}{8}\cos^{3}(\tilde{\phi})-\frac{1}{12}\cos^{4}(\tilde{\phi})\right)
\nonumber
\end{eqnarray}
\begin{eqnarray}
+\left(\frac{\delta}{b}\right)^{2}\left(\frac{25}{256}\cos(\tilde{\phi})-\frac{29}{256}\cos^{3}(\tilde{\phi})+\frac{1}{64}\cos^{5}(\tilde{\phi})\right)).
\end{eqnarray}
We set up $\tilde{\phi}=\frac{\pi}{2}+\tilde{\alpha}$ in equation (\ref{eq:Vfinal}), such  that 
$V\left(\frac{\pi}{2}+\tilde{\alpha}\right)=0$, and use the expansion:
\begin{eqnarray}
\sin(\tilde{\alpha})=\epsilon\chi_{1}+\epsilon^{2}\chi_{2}+\epsilon^{3}\chi_{3}+.....,
\end{eqnarray}
where the $\chi_{k}$ are constants to be determined. The result is to second order in $\epsilon$:
\begin{eqnarray}
\epsilon\left(-\chi_{1}+\frac{2}{3}\right)+\epsilon^{2}\left(-\chi_{2}+\frac{1}{3}\chi_{1}+\frac{\delta}{8b}\chi_{1}-\frac{4}{9}-\frac{\delta}{b}\right)+ ........=0
\end{eqnarray}
implying:
\begin{equation}
\chi_{1}=\frac{2}{3},
\end{equation}
and
\begin{equation}
\chi_{2}=-\frac{2}{9}-\frac{11}{12}\left(\frac{\delta}{b}\right).
\end{equation}
Thus, $\sin(\tilde{\alpha})$ is given to second order by
\begin{eqnarray}
\sin(\tilde{\alpha})=\frac{2}{3}\epsilon-\left(\frac{2}{9}+\frac{11}{12}\left(\frac{\delta}{b}\right)\right)\epsilon^{2}.
\end{eqnarray}
$\tilde{\alpha}$ is obtained from the last equation using the Taylor series of $\arcsin(x)$ around
$x=0$:
\begin{equation}
arcsin(x) = x +\frac{x^3}{6} + O(x^5)
\end{equation}
to obtain:
\begin{eqnarray}
\tilde{\alpha}=\frac{2}{3}\epsilon-\left(\frac{2}{9}+\frac{11}{12}\left(\frac{\delta}{b}\right)\right)\epsilon^{2}+....
\end{eqnarray}
From the definition of the strained coordinate $\tilde{\phi}$ in (\ref{eq:strained}) we have:
\begin{eqnarray}
\left(\frac{\pi}{2}+\alpha\right)=\frac{\left(\frac{\pi}{2}+\tilde{\alpha}\right)}{1+\frac{3}{8}\left(\frac{\delta}{b}\right)\epsilon+\left(-\frac{5}{12}-\frac{1}{4}\left(\frac{\delta}{b}\right)-\frac{21}{256}\left(\frac{\delta}{b}\right)^{2}\right)\epsilon^{2}+ .....}
\end{eqnarray}
From the last  equation, and using the Taylor expansion of $\frac{1}{1+x} = \sum_{n = 0}^{\infty} (-1)^{n}x^{n}$ around $x=0$, we get:
\begin{eqnarray}
\alpha=\epsilon\left(\frac{2}{3}-\frac{3\pi}{16}\left(\frac{\delta}{b}\right)\right)+\epsilon^{2}\left(\frac{5\pi}{24}-\frac{2}{9}+\left(\frac{\pi}{8}-\frac{7}{6}\right)
\left(\frac{\delta}{b}\right)+\frac{57\pi}{512}\left(\frac{\delta}{b}\right)^{2}\right).
\end{eqnarray}
The total deflection angle is then $\Omega=2\alpha$. Replacing the values of $\epsilon$ and $\delta$, we can finally write:

\begin{eqnarray}
\Omega=\frac{4GM}{bc^{2}}-\frac{3GQ^{2}}{16\epsilon_{0}b^{2}c^{4}}+\left(\frac{15\pi}{4}-4\right)\left(\frac{GM}{bc^{2}}\right)^{2}\nonumber
\label{eq:Omega}
\end{eqnarray}
\begin{eqnarray}
+\left(\frac{3}{8}-\frac{7}{2\pi}\right)\frac{G^{2}MQ^{2}}{\epsilon_{0}b^{3}c^{6}}
+\frac{57}{1024}\frac{G^{2}Q^{4}}{\pi\epsilon_{0}^{2}b^{4}c^{8}}.
\end{eqnarray}
\

When $Q=0$, this result matches the one obtained in \cite{Bodenner,Fischback,Richter,Epstein}.
A third-order solution for $\Omega(\epsilon)$ can be obtained from equation (\ref{eq:Vdif3}) following a procedure similar to the one used in this section. Note that getting a higher-order solution conserves the lower-order terms.

\section{Pad\'e Approximants}
The formal Taylor expansion of $\Omega$ around $\epsilon=0$ is:
\begin{eqnarray}
\Omega=\kappa_{1} \epsilon^{1}+\kappa_{2} \epsilon^{2}+\kappa_{3} \epsilon^{3}+.....
\label{eq:Omegakappa}
\end{eqnarray}

We can check easily that as $\epsilon \rightarrow 1$, $\Omega\left(\epsilon\right) \rightarrow \infty$.

As $b\rightarrow r_c$, the deflection angle must go to infinity ($\Omega\rightarrow\infty$). This is because the photons starts going into a closed orbit, the photon circular orbit. This implies that $\Omega(\epsilon)$ must diverge at $b=r_c$ ( in other words  $\Omega(\epsilon)$ has a singularity at  $\epsilon = 1$). The problem is that a polynomial does not have asymptotes. If we want to recover the asymptote, then we need a rational function \cite{MarinRodriguez}.

Rational functions of the form:

\begin{equation}
f^{[M/N]}(x)=\frac{p_o+p_1x+p_2x^2+...+p_Mx^M}{1+q_1x+q_2x^2+...q_Nx^N}
\end{equation}

are called the Pad\'e approximants of a function $f(x)$, if the equation

\begin{equation}
\frac{d^kf^{[M/N]}}{dx^{k}}=\frac{d^kf}{dx^k}
\end{equation}
is true for all $k\in[0,M+N]$ \cite{Pade1}.

Also, if $f(x)$ can be written as a power series:

\begin{equation}
f(x)=\sum_{k=0}^\infty c_kx^k \label{eq:taylor}
\end{equation}
then, the first $M+N$ terms of the Taylor expansion of $f^{[M/N]}$ around $x=0$ must match with the coefficients of equation (\ref{eq:taylor}).

To work with the expansion

\begin{equation}
\Omega=\frac{4GM}{bc^2}-\frac{3}{16}\frac{GQ^2}{\epsilon_o b^2 c^4}+\left(\frac{15\pi}{4}-4\right)\left(\frac{GM}{bc^2}\right)^2+\left(\frac{3}{8}-\frac{7}{2\pi}\right)\frac{G^2MQ^2}{\epsilon_ob^3c^6}+\frac{57}{1024\pi}\frac{G^2Q^4}{\epsilon_o^2b^4c^8}
\end{equation}
it is better to write it in terms of $\epsilon_1= \epsilon=\frac{3GM}{bc^2}$ and $\epsilon_2=\frac{GQ^2}{4\pi\varepsilon_o b^2 c^4}$. Using this relations, the Taylor expansion of $\Omega$ would be:

\begin{equation}
\Omega=\frac{4}{3}\epsilon_1-\frac{3\pi}{4}\epsilon_2+\left(\frac{5\pi}{12}-\frac{4}{9}\right)\epsilon_1^2+\left(\frac{\pi}{2}-\frac{14}{3}\right)\epsilon_1\epsilon_2+\frac{57\pi}{64}\epsilon_2^2 \label{eq:expansion}
\end{equation}

For example, using the second order Taylor polynomial of $\Omega$, one can recover the univariate $[1/1]$ Pad\'e approximant of $\Omega$, for $\epsilon_1=\epsilon$, and $\epsilon_2=0$ (as in \cite{MarinRodriguez}):

\begin{equation}
\Omega^{[1]}(\epsilon)=\frac{64\epsilon}{48+\left(16-15\pi\right)\epsilon}
\end{equation}

Now, if we want to calculate the first order Pad\'e approximants for equation (\ref{eq:expansion}), including the charge term $\epsilon_2$, we would need a multivariate expansion.

Working with Pad\'e approximants is very difficult in some cases, because there is not guarantee to obtain accurate results. However, some algorithms have been developed that facilitate the work \cite{Pade1,Pade2}. This is not the case for multivariate Pad\'e approximants. First, it is difficult to chose the correct numerator and denominator, because not for all cases the system of equations have exact solutions. Also, one can obtain inaccurate results. Second, there is no optimal algorithm that gives us the correct number of equations we need to find the coefficients of the expansion \cite{Pade2,Pade3}.

Nevertheless, in the present article we use the algorithm showed in \cite{Pade4}, and then we analyse the results to find the best fit. Suppose the function $f(x,y)$ can be decomposed in power series as:

\begin{equation}
f(x,y)=\sum_{i,j\geq 0}^\infty c_{ij} x^iy^j
\end{equation}

The Pad\'e multivariate problem consist in finding two polynomials of the form $p(x,y)=\sum_{k= 0}^{M} A_{k}(x,y)$ and $q(x,y)=\sum_{k= 0}^{N} B_{k}(x,y)$, where $A_k(x,y)$ and $B_k(x,y)$ are polynomials of order $k$; such that $p(x,y)$ and $q(x,y)$ satisfy:

\begin{equation}
f(x,y)=\frac{p(x,y)}{q(x,y)}
\end{equation}

Or equivalently:

\begin{equation}
q(x,y)\sum_{i,j\geq 0}^{M+N} c_{ij} x^iy^j-p(x,y)=0 \label{eq:poli}
\end{equation}

For simplicity, we define the polynomials $C_k(x,y)=\sum_{i+j=k} c_{ij} x^iy^j$. This polynomials are: $C_0=c_{00}$, $C_1(x,y)=c_{10}x+c_{01}y$, $C_2(x,y)=c_{20}x^2+c_{11}xy+c_{02}y^2$, etc. Then, form condition (\ref{eq:poli}) we can recover the following system of equations to find the polynomials $A_k$ and $B_k$:

\begin{equation}
\left\lbrace\begin{array}{c}
C_0(x,y)B_0(x,y)=A_0(x,y) \\
C_1(x,y)B_0(x,y)+C_0(x,y)B_1(x,y)=A_1(x,y) \\
\vdots \\
C_M(x,y)B_0(x,y)+\dots+C_{M-N}(x,y)B_N(x,y)=A_M(x,y)
\end{array}\right.
\label{eq:alg1} 
\end{equation}

\begin{equation}
\left\lbrace\begin{array}{c}
C_{M+1}(x,y)B_0(x,y)+...+C_{M-N+1}(x,y)B_N(x,y)=0 \\
C_{M+2}(x,y)B_0(x,y)+...+C_{M-N+2}(x,y)B_N(x,y)=0 \\
\vdots \\
C_{M+N}(x,y)B_0(x,y)+\dots+C_{M}(x,y)B_N(x,y)=0
\end{array}\right.
\label{eq:alg2}
\end{equation}

And, the approximant can be written as

\begin{equation}
f^{[M/N]}=\frac{\sum_{k= 0}^{M} A_{k}(x,y)}{\sum_{k= 0}^{N} B_{k}(x,y)}
\end{equation}

From (\ref{eq:expansion}), one can retrieve the coefficients: $C_0=0$, $C_1(x,y)=\frac{4}{3}\epsilon_1-\frac{3\pi}{4}\epsilon_2$, $C_2(x,y)=\left(\frac{5\pi}{12}-\frac{4}{9}\right)\epsilon_1^2+\left(\frac{\pi}{2}-\frac{14}{3}\right)\epsilon_1\epsilon_2+\frac{57\pi}{64}\epsilon_2^2$. The problem is that setting $C_0=0$ retrieves problems like divisions by zero. To avoid that we set $C_0=\pi/2$, and then we only need to subtract $\pi/2$ from the final expression.

To see how the algorithm must be used, lets set $M=0$ and $N=1$ to calculate $\Omega^{[0/1]}$:

\begin{equation}
\Omega^{[0/1]}=\frac{A_o}{B_o+B_1(\epsilon_1,\epsilon_2)}-\frac{\pi}{2}
\end{equation}

Here we are already subtracting the $\pi/2$ term. From equations (\ref{eq:alg1}) and (\ref{eq:alg2}), one can find the following system of equations:

\[ C_o B_o=A_o\]

\[ C_1+ C_o B_1=0\]

As it can be seen, with these equations we have one degree of freedom, so we will set $B_o=1$. Then:

\[A_o=\frac{\pi}{2}\]

\[B_1=-\frac{C_1}{C_o}=-\frac{\frac{4}{3}\epsilon_1-\frac{3\pi}{4}\epsilon_2}{\pi/2}\]

Replacing these expressions in the equation of $\Omega^{[0/1]}$, and simplifying we have: 

\begin{equation}
\Omega^{[0/1]}(\epsilon_1,\epsilon_2)=\frac{16\pi\epsilon_1-9\pi^2\epsilon_2}{12\pi- 32\epsilon_1+ 18\pi\epsilon_2} \label{eq: [0/1]}
\end{equation}

Using similar steps, we get the following first order Pad\'e approximants of $\Omega(\epsilon_1,\epsilon_2)$:

\begin{equation}
\Omega^{[0/2]}(\epsilon_1,\epsilon_2)=\frac{\frac{\pi^2}{3}\epsilon_1 - \frac{3\pi^3}{16}\epsilon_2-\left(\frac{8\pi}{9}+ \frac{\pi^2}{9}- \frac{5\pi^3}{48} \right)\epsilon_1^2-\left(\frac{\pi^2}{6}-\frac{\pi^3}{8}\right)\epsilon_1\epsilon_2 - \frac{15\pi^3}{256}\epsilon_2^2}{\frac{\pi^2}{4}-\frac{2\pi}{3} \epsilon_1 + \frac{3\pi^2}{8}\epsilon_2+ \left(\frac{16}{9}+\frac{2\pi}{9}-\frac{5\pi^2}{24} \right)\epsilon_1^2+\left(\frac{\pi}{3}-\frac{\pi^2}{4}\right)\epsilon_1\epsilon_2 + \frac{15\pi^2}{128}\epsilon_2^2} \label{eq: [0/2]}
\end{equation}

\begin{equation}
\Omega^{[1/1]}(\epsilon_1,\epsilon_2)=\frac{\left(\frac{4}{3}\epsilon_1-\frac{3}{4}\pi\epsilon_2\right)^2}{\frac{4}{3}\epsilon_1-\frac{3}{4}\pi\epsilon_2-\left(\frac{5\pi}{12}-\frac{4}{9}\right)\epsilon_1^2-\left(\frac{\pi}{2}-\frac{14}{3}\right)\epsilon_1\epsilon_2-\frac{57\pi}{64}\epsilon_2^2} \label{eq:[1/1]}
\end{equation}

In these results, the $\pi/2$ term was already subtracted. Now, we need to test this approximants to analyse what is the best fit for the numerical points.


\section{Analysis and numerical tests}

\begin{figure}[htpb]
  \centering
  \includegraphics[angle=0,width=0.6\textwidth,]{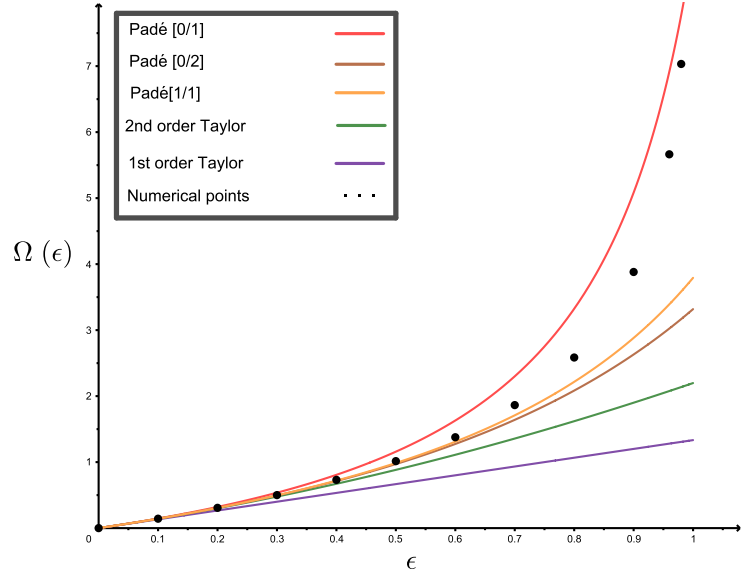}
  \caption{Plot of $\Omega(\epsilon)$ vs $\epsilon$ for $\epsilon'=0$.}
  \label{fig:fig1}
\end{figure}

\begin{figure}[htpb]
  \centering
  \includegraphics[angle=0,width=0.6\textwidth,]{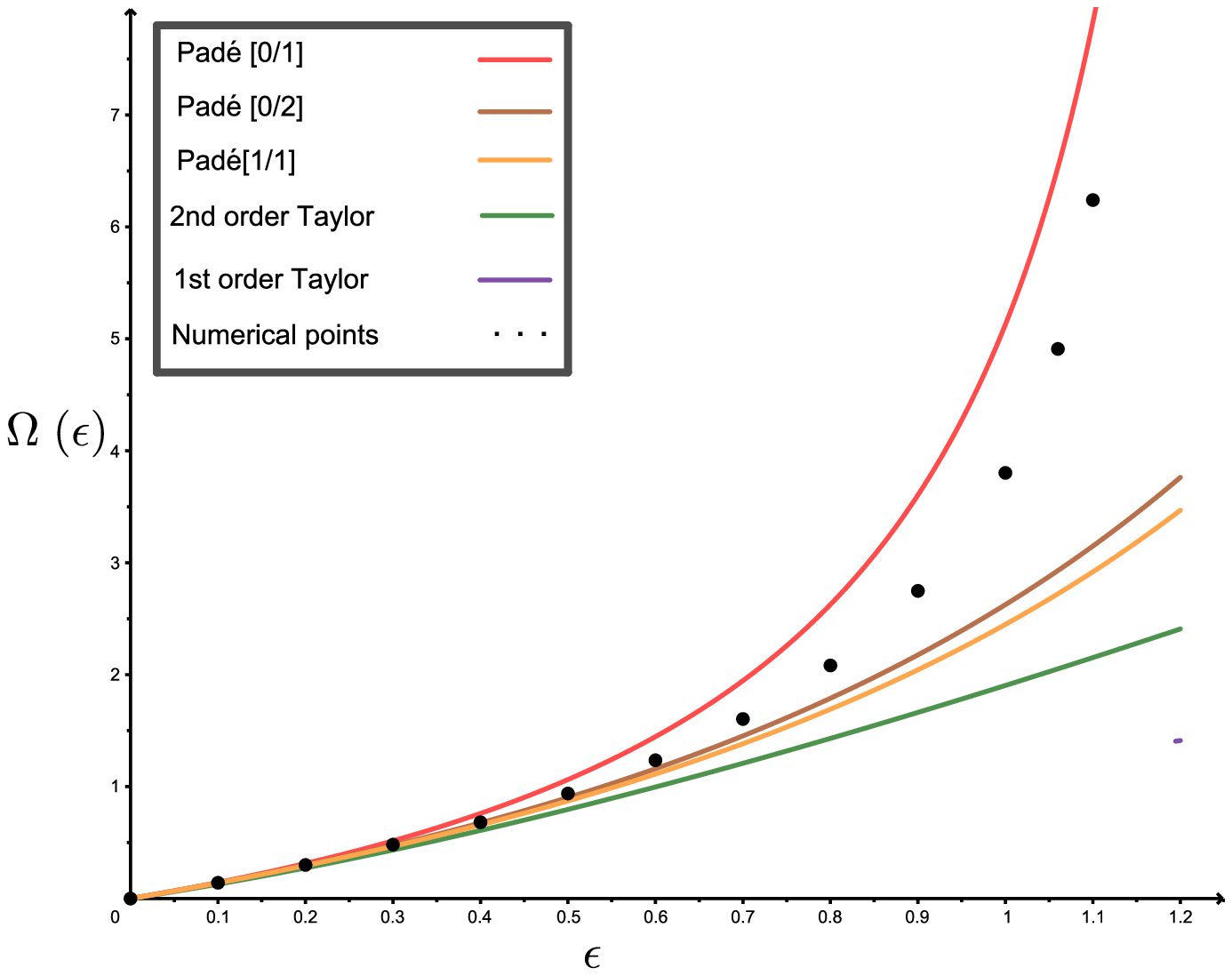}
  \caption{Plot of $\Omega(\epsilon)$ vs $\epsilon$ for $\epsilon'=\epsilon/18$.}
  \label{fig:fig2}
\end{figure}

\begin{figure}[htpb]
  \centering
  \includegraphics[angle=0,width=0.6\textwidth,]{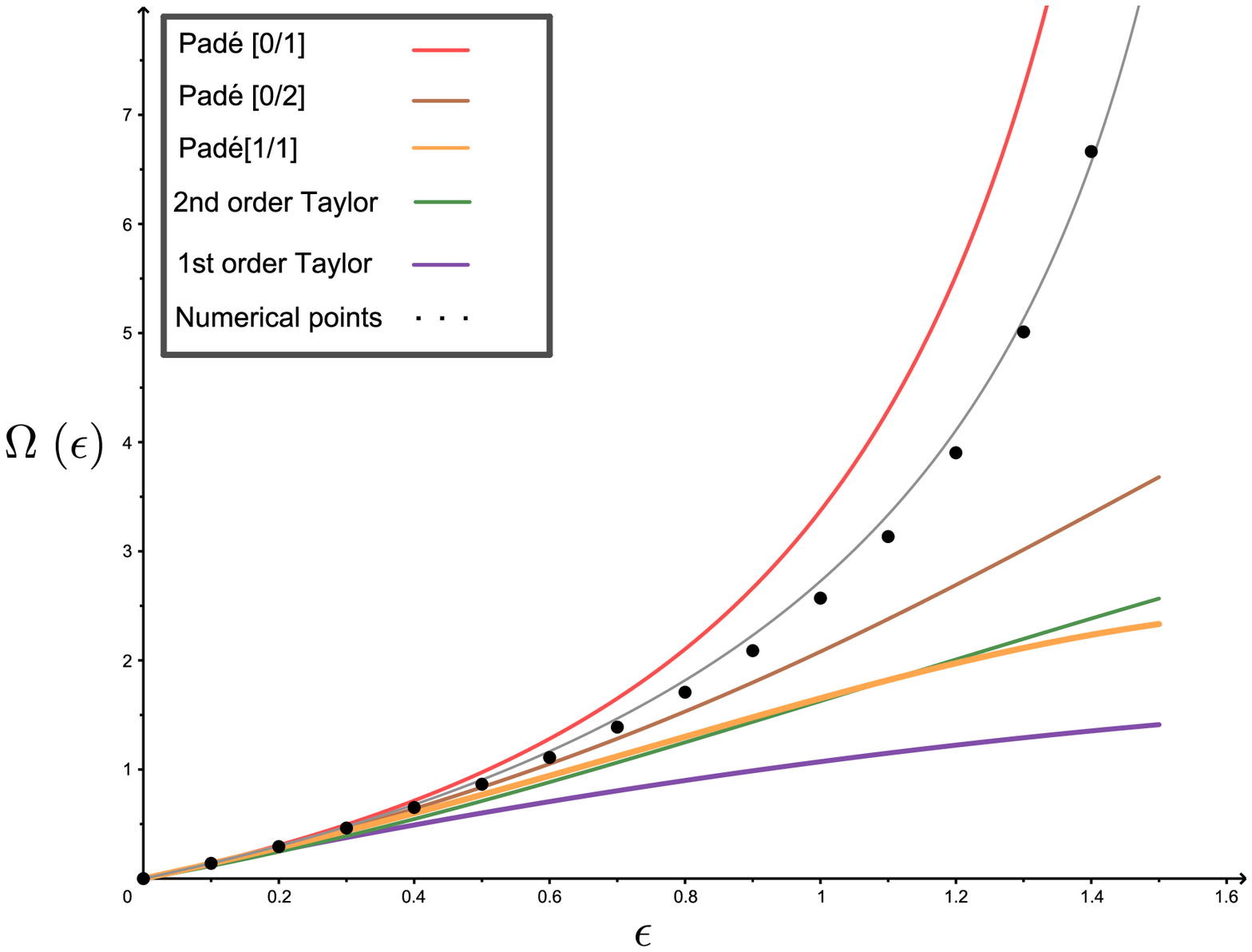}
  \caption{Plot of $\Omega(\epsilon)$ vs $\epsilon$ for $\epsilon'=\epsilon/9$.}
  \label{fig:fig3}
\end{figure}

Now we need to determine which Pad\'e approximant is the best fit for $\Omega$. In figures (\ref{fig:fig1}), (\ref{fig:fig2}), (\ref{fig:fig3}) are shown the plots of $\Omega$ in function of $\epsilon=\epsilon_1$ for the different approximations, including Taylor polynomials and Pad\'e approximants, and for different values of $\epsilon_2=\epsilon'$.

As it can be seen, no Pad\'e approximant seems to be a good fit for numerical values. This can be solved in the following way: As the numerical data are between the approximation [0/1] and the approximation [0/2], and also the approximation [0/2] does not have a singularity while the [0/1] has the singularity closest to the real one, we can approximate $\Omega$ with the average between [0/1] and [0/2] to find the best fit.

\begin{figure}[htpb]
  \centering
  \includegraphics[angle=0,width=0.6\textwidth,]{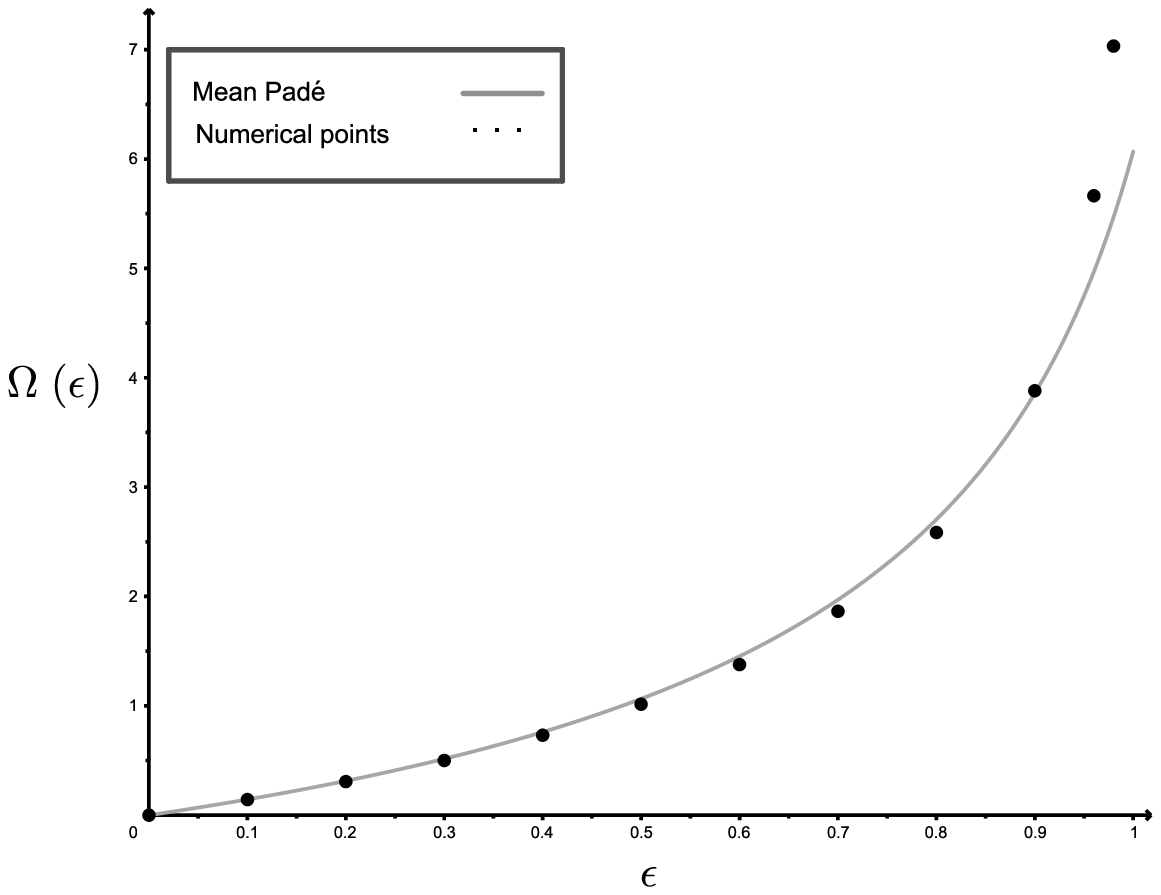}
  \caption{Plot of $\Omega(\epsilon)$ vs $\epsilon$ for $\epsilon'=0$.}
  \label{fig:fig4}
\end{figure}

\begin{figure}[htpb]
  \centering
  \includegraphics[angle=0,width=0.6\textwidth,]{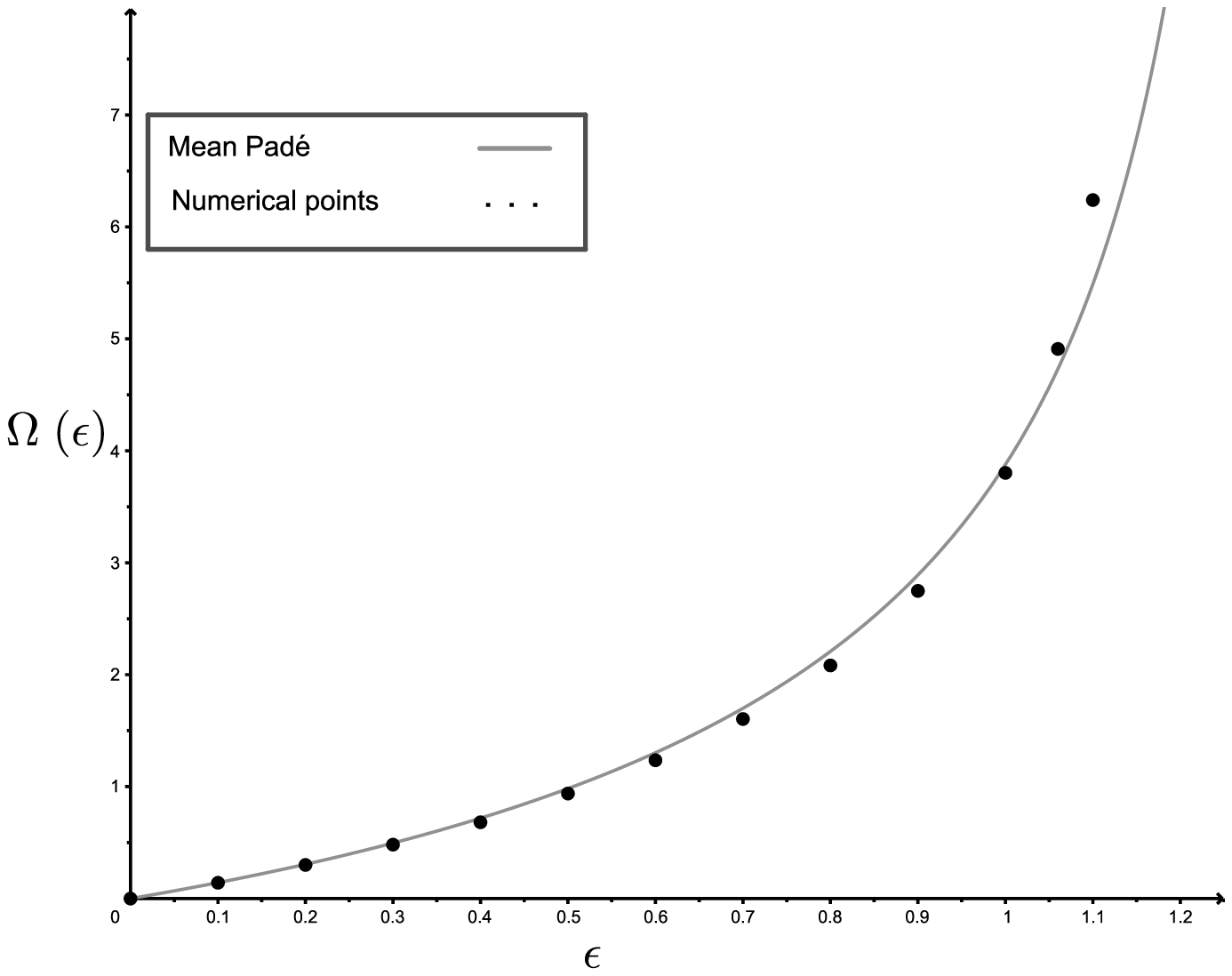}
  \caption{Plot of $\Omega(\epsilon)$ vs $\epsilon$ for $\epsilon'=\epsilon/18$.}
  \label{fig:fig5}
\end{figure}

\begin{figure}[htpb]
  \centering
  \includegraphics[angle=0,width=0.6\textwidth,]{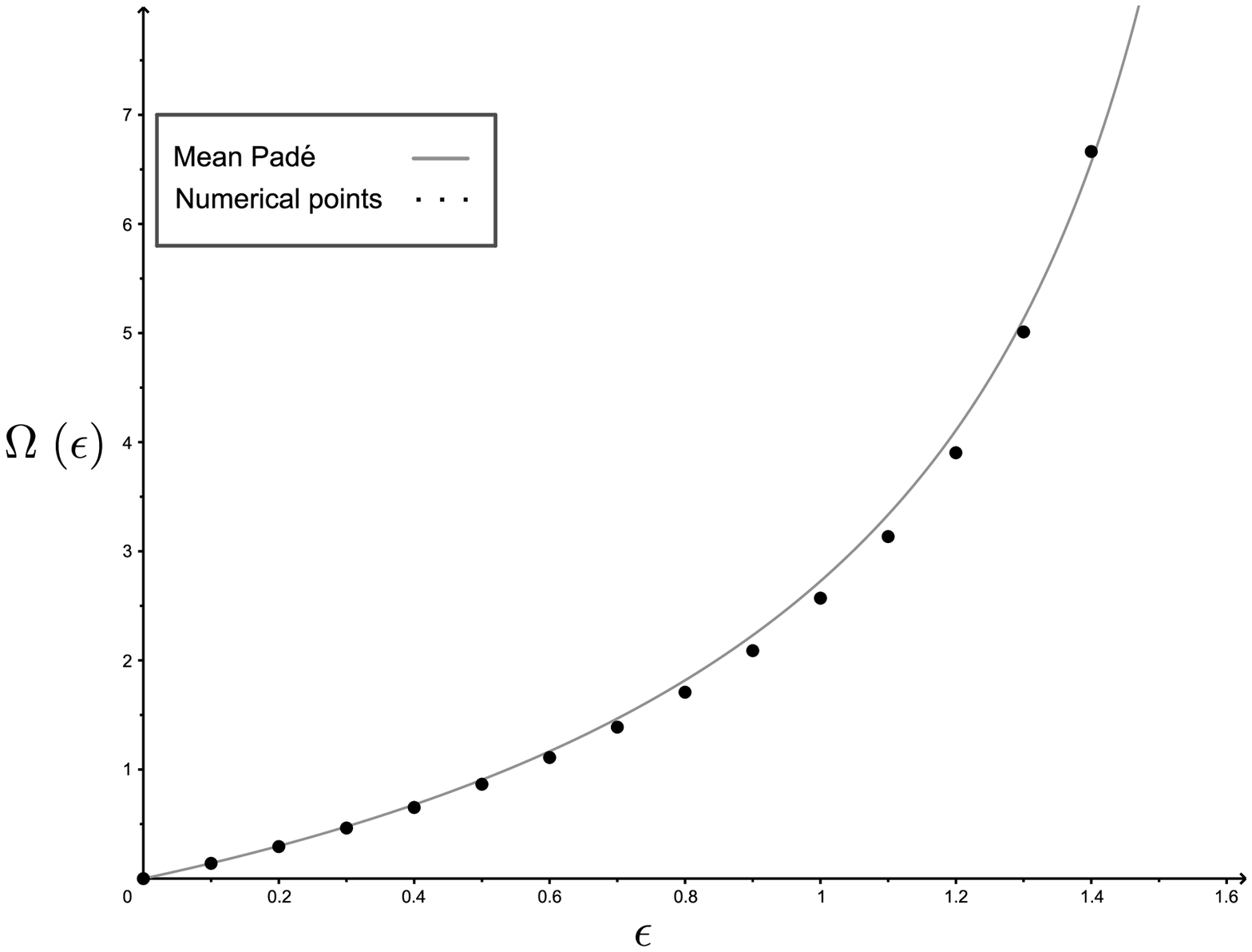}
  \caption{Plot of $\Omega(\epsilon)$ vs $\epsilon$ for $\epsilon'=\epsilon/9$.}
  \label{fig:fig6}
\end{figure}

In figures (\ref{fig:fig4}), (\ref{fig:fig5}), (\ref{fig:fig6}) are shown the plots of

\begin{equation}
\Omega = \frac{(\Omega^{[0/1]} + \Omega^{[0/2]})}{2}
\end{equation}
for different values of $\epsilon_2=\epsilon'$. As it can be seen, this term is a better fit of $\Omega$ for values of $r$ near to the singularity. Nevertheless, if it can be recovered more Taylor expansion terms, it is possible to calculate higher order Pad\'e approximants.

\section{Conclusions}
\label{Conclusions}

In this paper we have obtained  from the Reissner-Nordstr\"om metric the second-order terms for the deflection of light around a massive-charged black hole using the Lindstedt-Poincar\'e method to solve  the equation of motion of a photon around the compact object (equation (\ref{eq:Omega})). This method allows us to eliminate any and all secular terms that arise (like $\phi \sin(\phi)$) to get a well-behaved solution.  The corrections are performed developing the expansion in terms of $\epsilon=\frac{r_{c}}{b}$  and $\frac{\delta}{b}= \frac{Q^{2}}{6 \pi \epsilon_{0}Mc^{2} b}$. A third-order solution for $\Omega(\epsilon)$ (and higher order corrections) can be obtained from equation (\ref{eq:Vdif3}) following a procedure similar to the one used in section \ref{Second-order}.

 Also we have obtained multivariate Pad\'e approximants from the perturbation expansion. By analysing the different Pad\'e expansions, it was determined that the better fit for $\Omega$ is:

\begin{equation}
\Omega = \frac{(\Omega^{[0/1]} + \Omega^{[0/2]})}{2}
\end{equation}
were $\Omega^{[0/1]}$ and $\Omega^{[0/2]}$  are given by equations (\ref{eq: [0/1]}) and (\ref{eq: [0/2]}), respectively. This term is a better fit of $\Omega$ for values of $r$ near to the singularity (see figures (\ref{fig:fig4}), (\ref{fig:fig5}), (\ref{fig:fig6})). However, if we can recover  more Taylor expansion terms, it is possible to calculate higher order Pad\'e approximants.

It is important to mention that Pad\'e polynomials  were first  used in Cosmology with excellent results \cite{Christine,Alejandro,Salvatore}.

We are convinced  that this paper can be very useful for undergraduate and graduate students to learn the use of perturbative techniques as the Lindstedt-Poincar\'e method or the Pad\'e approximants for solving problems not only within the framework of the General Theory of Relativity,  but also in other fields of Physics.

\end{document}